\documentclass[12pt]{article}

\columnwidth\textwidth

\begin{document}

\newcommand{\be}{\begin{equation}}
\newcommand{\ee}{\end{equation}}
\newcommand{\beann}{\begin{eqnarray*}}
\newcommand{\eeann}{\end{eqnarray*}}
\newcommand{\bea}{\begin{eqnarray}}
\newcommand{\eea}{\end{eqnarray}}
\newcommand{\nn}{\nonumber}


\newcommand{\yman}{\ensuremath{\cal M}}
\newcommand{\xman}{\ensuremath{\Sigma}}
\newcommand{\rman}{\ensuremath{\rm I}\!{\rm R}}
\newcommand{\tman}{\ensuremath{\cal \tau}}


\newcommand{\D}{\ensuremath{\bf D}}
\newcommand{\G}{\ensuremath{\bf G}}
\newcommand{\Gab}{\ensuremath{{\bf G}_{\alpha\beta}}}
\newcommand{\Gmn}{\ensuremath{{\bf G}_{\mu\nu}}}
\newcommand{\iGab}{\ensuremath{{\bf G}^{\alpha\beta}}}
\newcommand{\iGmn}{\ensuremath{{\bf G}^{\mu\nu}}}
\newcommand{\Gam}{\ensuremath{{\bf G}_{\alpha\mu}}}
\newcommand{\Gan}{\ensuremath{{\bf G}_{\alpha\nu}}}
\newcommand{\Gbm}{\ensuremath{{\bf G}_{\beta\mu}}}
\newcommand{\Gbn}{\ensuremath{{\bf G}_{\beta\nu}}}
\newcommand{\iGam}{\ensuremath{{\bf G}^{\alpha\mu}}}
\newcommand{\iGan}{\ensuremath{{\bf G}^{\alpha\nu}}}
\newcommand{\iGbm}{\ensuremath{{\bf G}^{\beta\mu}}}
\newcommand{\iGbn}{\ensuremath{{\bf G}^{\beta\nu}}}

\newcommand{\T}{\ensuremath{\bf T}}
\newcommand{\X}{\ensuremath{\bf X}}
\newcommand{\Xj}{\ensuremath{\bf X}^j}
\newcommand{\Y}{\ensuremath{Y}}
\newcommand{\Ya}{\ensuremath{Y}^{\alpha}}
\newcommand{\Yb}{\ensuremath{Y}^{\beta}}
\newcommand{\Ym}{\ensuremath{Y}^{\mu}}
\newcommand{\Yn}{\ensuremath{Y}^{\nu}}

\newcommand{\N}{\ensuremath{\bf N}}
\newcommand{\Ni}{\ensuremath{{\bf N}^i}}
\newcommand{\g}{\ensuremath{\bf g}}

\newcommand{\M}{\ensuremath{\bf M}}
\newcommand{\Mab}{\ensuremath{{\bf M}^{\alpha\beta}}}
\newcommand{\Mmn}{\ensuremath{{\bf M}^{\mu\nu}}}
\newcommand{\Mam}{\ensuremath{{\bf M}^{\alpha\mu}}}
\newcommand{\Man}{\ensuremath{{\bf M}^{\alpha\nu}}}


\newcommand{\ya}{\ensuremath{Y^{\alpha}}}
\newcommand{\yb}{\ensuremath{Y^{\beta}}}

\newcommand{\gtt}{\ensuremath{G_{00}}}
\newcommand{\igtt}{\ensuremath{G^{00}}}
\newcommand{\gtx}{\ensuremath{G_{01}}}
\newcommand{\igtx}{\ensuremath{G^{01}}}
\newcommand{\gxx}{\ensuremath{G_{11}}}
\newcommand{\igxx}{\ensuremath{G^{11}}}

\begin{titlepage}

\noindent
\hspace*{11cm}  \\
\vspace*{0cm}
\begin{center}

{\Large Covariance and time regained in \\
canonical general relativity}

\vspace{0.5cm}

I. Kouletsis\\
February 2008 \\

\vspace{0.5cm}

\nopagebreak[4]

\begin{abstract}

Canonical vacuum gravity is expressed in generally-covariant form in
order that spacetime diffeomorphisms be represented within its
equal-time phase space. In accordance with the principle of general
covariance and ideas developed within history phase space formalisms
in Refs. \cite{kk}-\cite{s3}, the time mapping ${\T}: {\yman}
\rightarrow {\rman}$ and the space mapping ${\X}: {\yman}
\rightarrow {\xman}$ that define the Dirac-ADM foliation are
incorporated into the framework of the Hilbert variational
principle. The resulting canonical action encompasses all individual
Dirac-ADM actions, corresponding to different choices of foliating
vacuum spacetimes by spacelike hypersurfaces. The equal-time phase
space ${\cal P}=\{ g_{ij}, p^{ij}, Y^{\alpha}, P_{\alpha} \}$
includes the embeddings $Y^{\alpha}$ and their conjugate momenta
$P_{\alpha}$. It is constrained by eight first-class constraints.
The constraint surface $\cal C$ is determined by the
super-Hamiltonian and super-momentum constraints of vacuum gravity
and the vanishing of the embedding momenta. Deformations of the time
and space mappings, ${\delta}{\T}$ and ${\delta}{\X}$, and spacetime
diffeomorphisms, $V \in {\rm L}{\rm Diff}{\yman}$, induce symplectic
diffeomorphisms of ${\cal P}$. While the generator ${\cal
D}_{({\delta}{\T},{\delta}{\X})}$ of deformations depends on all
eight constraints, the generator ${\cal D}_{V}$ of spacetime
diffeomorphisms depends only on the embedding momentum constraints.
As a result, spacetime observables, namely, dynamical variables $F$
on $\cal P$ that are invariant under spacetime diffeomorphisms, $\{
F, {\cal D}_{V} \} {\mid}_{\cal C}=0$, are not necessarily invariant
under the deformations of the mappings, $\{ F, {\cal
D}_{({\delta}{\T},{\delta}{\X})} \} {\mid}_{\cal C} \neq 0$, nor are
they constants of the motion, $\{ F, \int d^3x \, {\cal H} \}
{\mid}_{\cal C} \neq 0$. Dirac observables form only a subset of
spacetime observables that are invariant under the transformations
of $\T$ and $\X$ and do not evolve in time. In this generally-covariant framework, the conventional interpretation of the canonical theory, due to Bergmann
and Dirac, amounts to postulating that the transformations of the
reference system $({\T},{\X})$ have no measurable consequences;
i.e., that {\it all} first-class constraints generate gauge transformations.
If this postulate is not deemed necessary, canonical gravity admits no classical problem of time.

\end{abstract}

\end{center}

\end{titlepage}

\section{Introduction}

\subsection{General covariance, determinism and the problem of evolution}

The variational principle for general relativity, with or without
sources, introduces a four-dimensional manifold $\yman$ and an
action functional $S[{\bf \Psi}]$ on $\yman$. The principle of
general covariance demands that all fields ${\bf \Psi}$ be subject
to variation in the action functional and satisfy
generally-covariant field equations. In the case of the vacuum
theory, where only the metric field is present on the spacetime
manifold, the set of solutions consists of all distinct metric
fields $\G$ on $\yman$ that satisfy the vacuum Einstein equations.
These solutions do not all correspond to physically distinct states
of the system. Considering that the manifold points are physically
indistinguishable prior to introducing the fields on $\yman$, any
two solutions that can be brought into coincidence by an element of
the group ${\rm Diff}{\yman}$ are regarded as representations of the
same physical state. The group ${\rm Diff}{\yman}$ is treated as the
gauge group of the theory, and each physical state is identified
with an equivalence class $\{{\G}\}$ of ${\rm Diff}{\yman}$-related
solutions on $\yman$. The set $\Gamma$ of all such equivalence
classes constitutes the set of physically distinct states of the
system.

In the canonical formalism, initiated by Dirac \cite{Dir} and
Arnowitt, Deser and Misner \cite{adm}, the same physical conclusion
can be drawn by considering the initial-value problem. General
relativity is not a deterministic dynamical system in the strict
sense. A characteristic of its canonical formulation is that a given
set of instantaneous data at an initial time $t_1$ may evolve, via
different choices of lapse and shift, to many different sets of such
data at a later time $t_2$. Nevertheless, a well-posed initial-value
problem arises if it is stipulated that all these sets of evolved
data characterise the same physical situation \cite{Berg}-\cite{Diracqm}. Within the framework of the Dirac-ADM phase space ${\cal P}=\{ (g_{ij}, p^{ij}) \}$, each set of permissible data $(g_{ij}(x), p^{ij}(x))$ on a given hypersurface
defines a point on the constraint surface ${\cal C} \subset {\cal
P}$, where $\cal C$ is determined by the first-class constraints.
All points in $\cal C$ to which an initial point can evolve via
arbitrary choices of lapse and shift lie in an orbit of the
Hamiltonian vector field generated by the first-class constraints.
The set $\Delta$ of such distinct orbits in $\cal C$, equipped with
an induced symplectic form, constitutes the so-called reduced phase
space of the theory. This set can be brought into a one-to-one
correspondence with the set $\Gamma$ of ${\rm Diff}{\yman}$-classes
of solutions on $\yman$ \cite{AbrMar}.

In this way, the original classification of physical states
according to the set $\Gamma$ is recovered, and the inability to
physically distinguish between evolved data in the canonical theory
may be attributed to the invariance of the spacetime action under
${\rm Diff}{\yman}$. In addition, the bijective correspondence
between the sets $\Gamma$ and $\Delta$ allows the physical
observables of the theory to be perceived either as functions on
$\Delta$, the so-called Dirac observables, or as functions on
$\Gamma$, which may be referred to as spacetime observables. In
either case, the physical observables remain invariant under the
dynamical evolution generated by the Hamiltonian, a fact which
implies that this evolution is not measurable. Only symmetries of
the reduced phase space, i.e., symplectic transformations of
$\Delta$, and equivalently of $\Gamma$, can be contemplated as being
measurable \cite{Hajicek}. Even if such symmetries are discovered in
general relativity, global obstructions are expected to arise in the
phase space \cite{Torre} which may prohibit such symmetries from
being interpreted as generators of the evolution of the system in
physical time. This problem of evolution may be regarded as the
classical core of the problem of time of quantum gravity.

\subsection{The missing representations of the group ${\rm Diff}{\yman}$}

Of particular relevance to the problem of evolution is the way in
which the group ${\rm Diff}{\yman}$ is considered to act on the
phase space of general relativity, and the connection between this
action and the dynamical evolution generated by the Hamiltonian. A
peculiar feature of the Dirac-ADM formalism is that, despite the
bijective correspondence between the sets $\Gamma$ and $\Delta$, the
${\rm Diff}{\yman}$-invariance of the spacetime action is reflected
only indirectly in the first-class constraints. More precisely,
although the canonical transformations generated by the Hamiltonian
can be linked to the diffeomorphisms of the spacetime manifold
$\yman$, the Lie algebra of ${\rm Diff}{\yman}$ cannot be mapped
onto the Poisson bracket algebra of the super-Hamiltonian and
super-momentum constraints. This inability to recover the action of
${\rm Diff}{\yman}$ directly within the conventional canonical
framework is not only noteworthy from the conceptual point of view,
but also contributes to the problems that hinder the canonical
quantisation of gravity.

The cause of this difficulty was diagnosed by Isham and Kucha\v{r}
\cite{ik}. The absence from the conventional phase space ${\cal P}=
\{ (g_{ij}, p^{ij}) \}$ of the embedding mappings $Y: \xman
\rightarrow \yman$ that connect the spacetime manifold $\yman$ with
the space manifold $\xman$ renders the direct canonical description of
spacetime objects impossible, and leads to the loss of the representations of ${\rm Diff}{\yman}$. In order to recover the action of ${\rm Diff}{\yman}$ within the canonical framework, this missing link must be re-established, and the gravitational configuration space must be extended by the space of embeddings from $\xman$ to $\yman$. This was achieved in Ref. \cite{ik} by
parameterising the Dirac-ADM action.

The process of parameterisation is tantamount to viewing the lapse
function and the shift vector as functionals of the embedding
mapping $Y: {\xman} \rightarrow {\yman}$, and then varying $Y$ in
the action. When applied to a generally-covariant system such as
general relativity, this procedure requires that four of the
components of the spacetime metric be limited by coordinate
conditions with respect to the foliation structure. The coordinate
conditions are needed in order that the lapse function and the shift
vector can indeed be regarded as functionals of the embedding
variable $Y$, and not as variables on their own. In addition, these
conditions ensure a well-posed initial-value problem. Without them,
the spacetime metric built by the canonical dynamical evolution
would be determined only up to a spacetime diffeomorphism
\cite{kuch86}.

As a result of limiting in Ref. \cite{ik} the spacetime metric by
the coordinate conditions, the original super-Hamiltonian and
super-momentum constraints get suspended, and new, modified,
constraints arise. In the resulting phase space $\{ (g_{ij}, p^{ij},
Y^{\alpha}, P_{\alpha}) \}$, augmented by the embeddings
$Y^{\alpha}(x)$ and their conjugate momenta $P_{\alpha}(x)$, a
direct correspondence between the spacetime and the canonical
descriptions emerges. This is attested via the construction of a
homomorphic mapping from the Lie algebra of ${\rm Diff}{\yman}$ into
the Poisson bracket algebra of the dynamical variables on the
extended phase space.

Viewed from the perspective of a variational principle, the
procedure of breaking the invariance of general relativity by
coordinate conditions and restoring it by parameterisation can be
associated with the coupling of gravity to matter fields. Kucha\v{r}
and Torre \cite{refflu1} derived the modified constraints of Isham
and Kucha\v{r} from an appropriate action functional, and recognised
the new terms as the energy-momentum density of a non-rotating,
heat-conducting, incoherent dust. Other coordinate conditions lead
to different constraint structures, some of which have been
investigated in Refs. \cite{refflu2}-\cite{bickuc}.

\subsection{Aim, motivation and main concept}

In this paper, a reformulation of the canonical method is considered, that permits the representation of the Lie algebra of ${\rm Diff}{\yman}$ within a suitable equal-time phase space for vacuum general relativity, without abandoning the standard constraints of this theory. The proposed formalism relies upon ideas and techniques that were developed in collaboration with K. Kucha\v{r} in Ref. \cite{kk} and yields results that are, in certain ways, parallel to the results of Savvidou \cite{s1}-\cite{s3}, derived within the context of the History Projection Operator\footnote{The Consistent Histories
interpretation of quantum theory was initiated by Griffiths \cite{g} and
developed by Omn\`{e}s \cite{o}-\cite{o4}, Gell-Mann and Hartle \cite{gh}-\cite{gh5}
and Isham \cite{isham}; see also \cite{il}-\cite{y}.} formalism for general relativity.

From a technical point of view, the only difference between the present formulation and the conventional formulation of Dirac and ADM is that the foliation is modelled as a variable, and is incorporated into the framework of the Hilbert variational principle. Such an approach is actuated by the desire to harmonise the canonical action with the principle of general covariance, and the recognition of the fact that, strictly speaking, this action is an extension of the Hilbert action. This is because, by construction, the canonical action requires a time foliation of $\yman$ by spacelike hypersurfaces to be introduced into general relativity as an additional geometric element.

Thereby, the notion of time is distinguished from that of a
spacetime coordinate and becomes dependent upon the spacetime metric $\G$.
Time is represented by a global scalar mapping $\T: \yman
\rightarrow {\rman}$ from the spacetime manifold $\yman$ to a
one-dimensional time manifold $\rman$ which has the topology of the
open line. The gradient ${\T}_{,\alpha}$ of this mapping is required to be
timelike with respect to $\G$. Accordingly, each choice of $\T$ represents a foliation of $\yman$ by spacelike hypersurfaces. On each
such hypersurface, the notion of space is represented by another
metric-dependent mapping ${\X}: {\yman} \rightarrow {\xman}$, whose
gradients ${\X}^i_{,\alpha}$ are required to be spacelike.

In order that the canonical theory be cast in generally-covariant form, all
fields upon which it is based must conform to the principle of general covariance. That is, they must be subject to variation in the action functional and satisfy generally-covariant field equations. In particular, field equations must be satisfied by $\T$ and $\X$. These equations must enforce the timelike and spacelike character of the gradients of these variables, but must otherwise
leave $\T$ and $\X$ undetermined in order to respect the arbitrariness of the spacelike foliation. As a result, a generally-covariant canonical action must necessary involve a greater number of non-dynamical variables than the Hilbert action. This causes the breaking of the bijective correspondence between its sets $\Gamma$ and $\Delta$ and, therefore, has repercussions for the functions defined on these sets; namely, the spacetime observables and the Dirac observables.

As it is evident, the breaking of this correspondence is a crucial property
of the covariant canonical action. In general, the sets $\Gamma$ and $\Delta$ reveal different aspects of a generally-covariant theory: on the one hand, the set $\Gamma$ is derived from the set of solutions by eliminating only the freedom associated with ${\rm Diff}{\yman}$. On the other hand, the set
$\Delta$ is derived from the set of solutions by eliminating the
freedom associated with the first-class constraints. In an arbitrary
generally-covariant framework, this latter freedom may be wider than
the former, because it depends upon the number of non-dynamical
variables present in the action; i.e., variables that are left
undetermined by the variational principle. In our particular case of
interest, after the non-dynamical variables $\T$ and $\X$ are
incorporated into the framework of the Hilbert variational
principle, the dynamical content of the resulting action, expressed
by the set $\Delta$, remains unaffected. However, the set $\Gamma$
of ${\rm Diff}{\yman}$-classes of solutions is extended by the
presence of these arbitrary fields. The set $\Delta$ becomes a
subset of $\Gamma$, and the Dirac observables form only a subset of
the spacetime observables. Since it is just this subset that weakly
commutes with the Hamiltonian, the evolution of the spacetime
observables is, in general, non-trivial.

In addition, the breaking of the bijective correspondence between
the sets $\Gamma$ and $\Delta$ is reflected within the equal-time
phase space $\cal P$ in the doubling of the first-class constraints.
Thus, it becomes possible in the covariant canonical formalism to
identify which constraints arise due to the diffeomorphism
invariance of the spacetime action and which arise due to the
non-dynamical character of the foliation. This lays the foundations
for, firstly, representing the Lie algebra of spacetime
diffeomorphisms by symplectic diffeomorphisms of $\cal P$ and,
secondly, for separating the canonical transformations generated
by spacetime diffeomorphisms from those generated by the
deformations of the foliation. In comparison, the bijection between
the sets $\Gamma$ and $\Delta$ in the Dirac-ADM formalism leads to
the entanglement of distinct concepts and the loss of general
covariance, while the preservation of this bijection via the
coordinate conditions causes the suspension of the vacuum
constraints in the covariant framework of Isham and Kucha\v{r}.

\subsection{The formalism}

A general procedure for incorporating the foliation into the
variational principle of a generally-covariant theory was developed
in Ref. \cite{kk}. It was designed originally for the purpose of
representing spacetime diffeomorphisms in the history phase space of
an arbitrary generally-covariant system, modelled in Ref. \cite{kk}
by the Bosonic string. This procedure respects the distinction
between the sets $\Gamma$ and $\Delta$, so it only needs to be
adapted to the circumstances of the gravitational theory. Thus, in
the same spirit, a time variable ${\T} :{\yman} \rightarrow {\rman}$
and a space variable ${\X} :{\yman} \rightarrow {\xman}$ are
incorporated into the Hilbert action as additional variable fields.
While the mapping ${\T}$ describes a slicing of the spacetime
manifold by spacelike hypersurfaces, namely, a time foliation, the
mapping ${\X}$ describes a congruence of timelike reference
world-lines, namely, a reference frame. The product mapping ${\T}
\times {\X}: {\yman} \rightarrow {\rman} \times {\xman}$ is inverse
to the foliation mapping $Y : {\rman} {\times} {\xman} \rightarrow
{\yman}$.

The variables $\T$ and $\X$ are coupled to the spacetime metric
$\G$. This coupling preserves the vacuum Einstein equations and also
ensures that, at the level of the solutions, the time foliation is
spacelike, and the reference frame is timelike, with respect to
$\G$. Apart from these essential restrictions, the variables $\T$
and $\X$ are left undetermined by the variational principle in order
to comply with the arbitrariness of the foliation. The resulting set
of solutions $\{ {\G}_{\alpha\beta}, {\T}, {\X}^i \}$ incorporates
the content of all individual Dirac-ADM actions in the sense that it
includes all causal reference systems that can be associated with
each vacuum spacetime. All the fields in this extended action
functional transform covariantly under the diffeomorphisms of
${\yman}$, so the general covariance of the formalism remains
manifest.

As in Ref. \cite{kk}, the transition from the spacetime action to
its Lagrangian counterpart on ${\rman} \times {\xman}$ is conceived
as a one-to-one transformation from the set of spacetime variables
$\{ {\G}_{\alpha\beta},{\T},{\X}^i \}$ to the set
$\{g_{ij},N,N^i,Y^{\alpha}\}$ of induced variables on ${\rman}
\times {\xman}$. This transformation is followed by a Legendre
transformation, which involves the foliation field $Y$. Since the
spacetime and the canonical frameworks remain interlinked, all
symmetries of the original field equations on $\yman$ are
transferred to the canonical theory. This provides the basis for
studying the transformations induced on the canonical fields
$\{g_{ij},p^{ij},N,N^i\}$ by the diffeomorphisms of $\yman$, as well
as by the deformations of the mappings $\T$ and $\X$.

Resembling the covariant formulation of Isham and Kucha\v{r}, the equal-time phase space ${\cal P}=\{(g_{ij},p^{ij},Y^{\alpha},P_{\alpha})\}$ includes the embeddings $Y^{\alpha}(x)$ and their conjugate momenta $P_{\alpha}(x)$. However, it is now constrained by eight first-class constraints. The constraint surface $\cal C$ is determined by the standard super-Hamiltonian and super-momentum constraints of vacuum gravity, $H=0$ and $H_i=0$, and the vanishing of the embedding momenta, $P_{\alpha}=0$. The Hamiltonian $\int d^3x \, {\cal H}$ is a linear functional of these eight first-class constraints, ${\cal
H}:= N H + N^i H_i + {\Lambda}^{\alpha} P_{\alpha}$.

\subsection{Summary of results}

The Hamiltonian $\int d^3x \, {\cal H}$ is regarded as the generator
of solutions rather than of symmetries; that is, its primary role is
considered to be the creation of solutions from permissible
instantaneous data. Symmetries of the field equations then act on
these solutions. Symmetries are generated by infinitesimal
transformations of the field variables that preserve the
linearisation of the field equations provided that these equations
hold. Each symmetry defines a mapping of solutions to solutions. Key
symmetries are induced by the diffeomorphisms of the manifolds
$\yman$ and the transformations of the mappings $\T$ and $\X$.

The diffeomorphisms of $\yman$ do not act on solutions in the same
way as the transformations of $\T$ and $\X$ do. Under the action of
${\rm Diff}{\yman}$, the spacetime metric $\G$ and the mappings $\T$
and $\X$ transform covariantly. This implies, in particular, that
the spacelike character of the time foliation and the timelike
character of the reference frame are respected. The foliation
variable $Y$, being inverse to ${\T} \times {\X}$, is transformed
arbitrarily by ${\rm Diff}{\yman}$, but the fields $g$, $p$, $N$ and
$N^i$ remain unchanged. In contrast, under the transformations of
the mappings $\T$ and $\X$, the spacetime metric $\G$ is left, by
definition, unchanged, but the fields $g$, $p$, $N$, $N^i$ and $Y$
are all transformed.

Special kinds of transformations of $\T$ and $\X$ are induced by the
diffeomorphisms of the manifolds $\rman$ and $\xman$. These move
individual hypersurfaces and individual worldlines, but they keep
the time foliation and the reference frame fixed; i.e., the final
collection of hypersurfaces and worldlines is the same as the
original. On account of this, the spacelike character of the time
foliation and the timelike character of the reference frame, with
respect to the unchanged $\G$, are preserved. This is not the case
for more general transformations of $\T$ and $\X$, unless these
transformations are allowed to depend fully upon solutions. Then it
is indeed possible to consider generalised symmetries
$\delta{\T}[{\G},{\T},{\X}]$ and $\delta{\X}[{\G},{\T},{\X}]$ that
sustain the compatibility between the mappings $\T$, $\X$ and the
unchanged $\G$.

Within the framework of the extended phase space $\cal P$, solutions
are visualised as curves lying in the subspace ${\rman} \times {\cal
C}$ of ${\rman} \times {\cal P}$. Symmetries of the field equations
are acting on these curves. The special transformations induced on
${\rman} \times {\cal P}$ by infinitesimal time diffeomorphisms $w
\in {\rm L}{\rm Diff}{\rman}$ and infinitesimal space
diffeomorphisms $u \in {\rm L}{\rm Diff}{\xman}$ are generated,
respectively, by the dynamical variables ${\cal D}_{w} = -\int d^3x
\, w {\cal H}$ and ${\cal D}_{u}=-\int d^3x \, u^i (H_i +
P_{\alpha}Y^{\alpha}_{,i})$. The more general transformations
induced on ${\rman} \times {\cal P}$ by the symmetries
$\delta{\T}[{\G},{\T},{\X}]$ and $\delta{\X}[{\G},{\T},{\X}]$ are
generated by the functional ${\cal D}_{({\delta}{\T},{\delta}{\X})}
=-\int d^3x \, \Big( \delta{\T} \, {\cal H} - \delta{\X}^i \, (H_i +
P_{\alpha}Y^{\alpha}_{,i}) \Big)$. This reduces to the generator
${\cal D}_{w}$ in the case where $\delta{\T}=w({\T})$ and
$\delta{\X}=0$, and to the generator ${\cal D}_{u}$ in the case
where $\delta{\T}=0$ and $\delta{\X}=u({\X})$. Analogous functionals
can be constructed within the Dirac-ADM phase space
$\{g_{ij},p^{ij}\}$.

On the other hand, the symmetries induced on ${\rman} \times {\cal
P}$ by infinitesimal spacetime diffeomorphisms $V \in {\rm L}{\rm
Diff}{\yman}$ are generated by a dynamical variable that has no
counterpart in the conventional phase space. This is the variable
${\cal D}_{V}= \int d^3x P_{\alpha}V^{\alpha}(Y)$, which depends
solely on the embedding variables and the vector field $V$. This
functional provides an anti-homomorphic mapping of vector fields in
the Lie algebra ${\rm L}{\rm Diff}{\yman}$ into the Poisson bracket
algebra on the phase space $\cal P$; i.e., a representation of
spacetime diffeomorphisms by symplectic diffeomorphisms of the phase
space.

The structure of the generators ${\cal
D}_{({\delta}{\T},{\delta}{\X})}$ and ${\cal D}_{V}$ reveals two
facts about canonical general relativity that lay unexpressed within
the conventional canonical formalism. First, the general covariance
of the theory is not reflected in the super-Hamiltonian and
super-momentum constraints but, instead, in the embedding momentum
constraints. Second, the orbits of the generators ${\cal D}_{V}$ and
$\int \,d^3x {\cal H}$ on the phase space $\cal P$ are distinct, in
accordance with the set $\Delta$ being a subset of $\Gamma$. This
eliminates any possibility of identifying the Hamiltonian functional
$\int d^3x \, {\cal H}$ with the generator of spacetime
diffeomorphisms, in agreement with Kucha\v{r}'s analysis of this
issue in Ref. \cite{kuch92}.

Although this distinct role of the Hamiltonian, as opposed to the
role of spacetime diffeomorphisms, cannot find an unambiguous
mathematical expression within the standard phase space
$\{(g_{ij},p^{ij})\}$ of vacuum gravity, it has been enacted in
formulations based on history phase spaces. In Ref. \cite{s1}, history representations of both the Lie algebra of ${\rm Diff}{\yman}$ and the Dirac algebra of the constraints are constructed within the context of the History Projection Operator formalism for general relativity. The foliation is introduced as a parameter in the formalism and satisfies an equivariance
condition \cite{s2}-\cite{s3}. The invariance of the canonical
action under ${\rm Diff}{\yman}$ was thereby established, and the
connection between this fact and the problem of time was studied.
The issue of the history quantisation of a spacelike foliation was
also analysed---see Ref. \cite{is}.

Alternative history representations of ${\rm Diff}{\yman}$ were
constructed in Ref. \cite{kk} in the context of the history phase
space of the Bosonic string. The equal-time formalism considered
here has inherited several features from that history formalism;
among them, the incorporation of the mappings $\T$ and $\X$ in the
variational principle, which makes the correspondence between the
sets $\Gamma$ and $\Delta$ many-to-one. This leads to the enrichment
of the notion of instantaneous observables and calls for the
revision of their dynamical evolution. As anticipated, two kinds of
observables arise on the equal-time phase space $\cal P$ of the
covariant canonical action: spacetime observables and Dirac
observables.

The spacetime observables are dynamical variables $F$ on ${\cal P}$
that commute on the constraint surface $\cal C$ with the generator
of spacetime diffeomorphisms, $\{ F, {\cal D}_{V} \} {\mid}_{\cal
C}=0$. While such functionals weakly commute with the embedding
momentum constraints, they do not necessarily weakly commute with
the super-Hamiltonian and super-momentum constraints. As a result,
they are not necessarily invariant under the deformations of the
mappings, $\{ F, {\cal D}_{({\delta}{\T},{\delta}{\X})} \}{\mid}_{\cal C} \neq 0$, nor are they constants of the motion, $\{ F, \int d^3x \, {\cal H} \} {\mid}_{\cal C} \neq 0$. On the other hand, the Dirac observables weakly commute with all eight first-class constraints, and hence also with $\int \, d^3x {\cal H}$. These are invariant under both the diffeomorphisms of $\yman$ and the transformations of the mappings $\T$ and $\X$, and form a subset of spacetime observables that remain frozen in
time. While the spacetime observables induce functions on $\Gamma$,
the Dirac observables induce functions on $\Delta$.

\subsection{Interpretation}

Regarded as an action functional on the spacetime manifold $\yman$,
the covariant canonical action is equivalent to the Hilbert action
coupled to causal reference systems $({\T},{\X})$. Although the
presence of these systems does not preclude the conventional
interpretation of vacuum gravity based upon Hilbert action, it does
imply that an additional postulate is necessary if this
interpretation is to be recovered within the framework of the
extended action. More precisely, the covariant canonical formalism
accepts two different interpretations, depending on whether physical
importance is ascribed to the entire set $\Gamma$ or solely to its
subset $\Delta \subset \Gamma$.

The second option amounts to the requirement, due to Bergmann
\cite{Berg} and Dirac \cite{Diracqm}, that all first-class
constraints generate gauge transformations. According to this
position, spacetime diffeomorphisms and deformations of the
mappings $\T$ and $\X$ have no measurable consequences. The mappings $\T$ and $\X$ are deemed unimportant, and the physical observables
coincide with the Dirac observables which are independent of these
mappings. Since the Dirac observables do not evolve in time, the
problem of evolution resurfaces in its standard form, as discussed
in the literature \cite{time1}-\cite{time8}. In this case, the
recovery of the representations of ${\rm Diff}{\yman}$ in the phase
space of the covariant canonical action is devoid of physical
significance.

Needless to say, prominence is given to the first option. According
to this position, the set $\Delta$ does not exhaust the observable
aspects of the theory. Significance is attributed to the entire set
$\Gamma$, and the selection of the mappings $\T$ and $\X$ as
additional variables advocates a specific physical proposition. This
concerns the issue of what constitutes a physical spacetime in
vacuum gravity; a long-standing issue that goes back to the founders
of general relativity: Hilbert formalised the notion that the
reference system in general relativity should be visualised as a
fluid which carries clocks that keep a causal time \cite{Hil}, and
Einstein used a similar idealisation in his book \cite{Ein}. Stachel
analysed the issue of observability in general relativity in Ref.
\cite{Sta}, and Rovelli introduced the so-called localised and
non-localised points of view in Ref. \cite{Rov}.

The concept of the reference fluid is realised in a mathematically
precise way by the mappings $\T$ and $\X$. These mappings bridge the
gap between observers and the system under observation in the
absence of a physical process of measurement. Observers are assumed
not to influence the gravitational system under observation.
Although their trajectories have to be timelike, they do not form
part of the physical system in the strict sense. Accordingly, the
interaction between the mappings $\T$, $\X$ and the metric $\G$ is
extremely tenuous. There is just enough interaction to distinguish
between the points of $\yman$, but not enough to disturb the
geometry. This is captured by the vanishing energy-momentum of the
fields $\T$ and $\X$ and the subsequent preservation of the vacuum
constraints in the canonical theory.

Regarding determinism, initial data do not uniquely determine the
evolution derived from the covariant canonical action, even after
the orbits of ${\rm Diff}{\yman}$ have been eliminated. There is
still freedom remaining in the theory due to the arbitrariness of
the foliation. However, this does not mean that the gravitational
system under observation has more freedom to evolve than it had
before; i.e., when it was described by Hilbert's action. The freedom
captured by the extended set $\Gamma$ only refers to the
possibilities of observation associated with a given physical state
$\delta \in \Delta$. As we shall see later, there is a whole set of
states $\{\gamma\}$ in $\Gamma$ associated with each physical state
$\delta \in \Delta$, all of which are ${\rm Diff}{\yman}$-invariant
but foliation-dependent.

Provided that the set $\Gamma$ is considered meaningful, each such
state $\gamma$ in the class $\{ \gamma \}$ is accepted as a distinct
measurable state of the physical state $\delta$. The underlying
assumption is that distinct measurements of a given physical
situation remain distinct even in the limit where the physical
interaction between the observers and the gravitational system
becomes negligible. In contrast, this kind of observability is
rejected in the formulation based on Hilbert's action. The focus is
placed there on the physical aspects of vacuum gravity in the strict
sense, and hence only on the gravitational field. Indicative of this
is the absence from that action functional of any variables
representing systems of reference. Instead, this concept is
relegated to the manifold charts or to external, auxiliary, elements
of the theory.

Systems of reference appear in the conventional framework as a
useful approximation, or even a necessary inconsistency, which is
eliminated from the physical interpretation of the theory. Observers
must have energy and momentum; otherwise, they cannot be considered
as being part of the system. Non-dynamical is interpreted as
physically unimportant, a fact which is declared by the bijective
correspondence between the sets $\Gamma$ and $\Delta$. In
particular, the metric field $\G$ at a given spacetime point is not
observable; at least, not according to the physical premises of the
theory as these follow from the selection of $\G$ as the sole
variable in Hilbert's action.

These premises change after the mappings $\T$ and $\X$ are adjoined
to the Hilbert action as additional variables. The emphasis is now
placed on the admission of arbitrary reference systems which
provides the empirical basis of Einstein's theory. These systems are
treated in the same way as the metric field is, and the set $\Gamma$
is extended. Spacetime points are individuated by the presence of
both the metric $\G$ and the fields $\T$ and $\X$, and the
interactions $g$, $N$, $N^i$ between these fields are ${\rm
Diff}{\yman}$-invariant and hence measurable. Within the extended
phase space of the covariant canonical action, the ${\rm
Diff}{\yman}$-induced first-class constraints generate gauge
transformations, but the deformations of the mappings $\T$ and $\X$,
despite being first-class, generate measurable changes of the time
foliation and the reference frame. The aspects of the problem of
time that touch on the classical theory are in this way overcome.

\subsection{Presentation}

The presentation is organised as follows. Section 2 summarises the
relevant aspects of canonical general relativity, with particular
emphasis being placed on the set $\Gamma$ of ${\rm
Diff}{\yman}$-classes of solutions and the set $\Delta$ of
first-class orbits. Section 3 illustrates the extension of the
Hilbert action by non-dynamical variables and explains the
subtleties of this procedure. This sets the stage for the main
technical part of the paper, which begins in section 4. There, the
extended action is introduced in its spacetime form, and the
corresponding canonical theory is derived. Section 5 investigates
the set of solutions and the sets $\Gamma$ and $\Delta$ of the
covariant canonical action and compares them with the corresponding
sets of the Hilbert action. Section 6 describes the symmetries
induced on the solutions of the field equations by the
diffeomorphisms of $\yman$ and the deformations of the mappings $\T$
and $\X$. Section 7 is concerned with the extended equal-time phase
space $\cal P$, the action of symmetries on the instantaneous data,
and their representations by symplectic diffeomorphisms of $\cal P$.
Finally, section 8 considers the spacetime observables and their
dynamical evolution, and discusses some conceptual aspects of the
problem of time that are elucidated by the proposed formalism.

\section{Background}

This section contains a summary of those aspects of general relativity
that are pertinent to this paper. It is a collection of standard results. Conventions and terminology are introduced, and simplifying assumptions are made when necessary.

\subsection{The set $\Gamma$ of ${\rm Diff}{\yman}$-classes of solutions}

The action principle for general relativity postulates a
four-dimensional background manifold $\yman$ and an action
functional $S[{\bf \Psi}]$ on $\yman$. The variables ${\bf \Psi}$
include a metric field $\G$ and possibly other geometrical objects.
The set of all kinematically admissible configurations of ${\bf
\Psi}$ will be denoted by ${\rm Virt}{\yman}$ and referred to as
{\it the set of virtual fields}. For example, in vacuum gravity,
${\bf \Psi}$ consists only of the metric field $\G$, and the set
${\rm Virt}{\yman}$ becomes the set ${\rm Riem}{\yman}$ of all
pseudo-Riemannian metrics on $\yman$. Each manifold
\begin{equation}
{\yman}_{\bf \Psi} := \big( {\yman}, {\bf \Psi} \big) \; ,
\label{virtual manifold}
\end{equation}
associated with a configuration ${\bf \Psi} \in {\rm Virt}{\yman}$,
will be referred to as a {\it virtual manifold}. After the variation
of the action functional, the set ${\rm Sol}{\yman} \subset {\rm
Virt}{\yman}$ of solutions consists of all ${\bf \Psi}$ in ${\rm
Virt}{\yman}$ that satisfy Einstein's equations. If ${\bf \Psi}$
belongs to ${\rm Sol}{\yman}$, ${\yman}_{\bf \Psi}$ will be called a
{\it solution manifold}.

There are as many distinct solution manifolds as there are distinct
${\bf \Psi} \in {\rm Sol}{\yman}$. However, not all of them
represent physically distinct states of the system. If there exists
a diffeomorphism $D: {\yman} \rightarrow {\yman}$ such that
\begin{equation}
{\bf \Psi}_2=D_{*}{\bf \Psi}_1 \, , \label{diffm-invariance}
\end{equation}
where $D_{*}$ is the push-forward mapping, the solution manifolds
${\yman}_{{\bf \Psi}_1}$ and ${\yman}_{{\bf \Psi}_2}$ are considered
equivalent; i.e., representations of the same physical state. For
asymptotically flat spacetimes, the group ${\rm Diff}{\yman}$ has to
be restricted so that it includes only those diffeomorphisms that
act trivially at `spatial infinity'. However, for simplicity,
$\yman$ will be assumed spatially compact in this paper.

The above considerations imply that the solution manifolds are
divided into equivalence classes. Each equivalence class of
solutions in ${\rm Sol}{\yman}$ is identified with an element
$\gamma$ of the quotient set
\begin{equation}
{\Gamma}:={\rm Sol}{\cal M}/o \, , \label{Gamma}
\end{equation}
where $o$ denotes the orbits of ${\rm Diff}{\yman}$ in ${\rm
Sol}{\yman}$. The set $\Gamma$ will be referred to as {\it the set
of ${\rm Diff}{\yman}$-classes of solutions}. The points of the
background manifold ${\yman}$ get entangled in $\Gamma$ with the
solutions ${\bf \Psi}$. This underlines the position that the points
of the background $\yman$ are physically indistinguishable in the
absence of fields. In other words, the set $\Gamma$ captures the
diffeomorphism invariance and background independence of general
relativity. These two properties will be jointly referred to as {\it
general covariance}.

\subsection{The time, space, and foliation mappings}

The manifold ${\yman}$ is assumed to be globally hyperbolic in order
that Cauchy surfaces exist. By a theorem of Geroch \cite{Geroch},
$\yman$ has the topology ${\Sigma} \times {\rm I}\!{\rm R}$.
Following Ref. \cite{kk}, the elements $y$ of $\yman$ will be called
{\it events}, the elements $x$ of $\xman$ will be called {\it
points}, and the elements $t$ of $\rman$ will be called {\it
moments}. Their coordinate representations are respectively
$y^{\alpha}$, $x^i$ and $t$. A global time mapping $\T$ is a
function
\begin{equation}
{\T} \, : \, {\yman} \, \rightarrow \, {\rman} \, \, \, \, {\rm by}
\, \, \, \, y \in {\yman} \, \mapsto \, t = {\T}(y) \in {\rman}
\label{eq:intro of T}
\end{equation}
from $\yman$ to the real numbers. For each solution manifold
${\yman}_{{\bf \Psi}}$, the gradient ${\T}_{,\alpha}$ of $\T$ has to
be timelike with respect to the metric field $G \in {\bf \Psi}$;
i.e., ${\iGab} {\T}_{,\alpha} {\T}_{,\beta} < 0$. This means that $\T$ must depend on ${\bf \Psi}$. Each such mapping $\T$ associates a spacelike hypersurface ${\Sigma}^{\T}_{(t)}$ in $\yman$ with a moment $t$ of $\rman$:
\begin{equation}
{\Sigma}^{\T}_{(t)} = \Big\{ y {\in} {\yman} : {\T}(y) = t {\in}
{\rman} \Big\} \; . \label{eq:sigma}
\end{equation}
Such a hypersurface will be called {\it an instant} and their
collection
\begin{equation}
{\Sigma}^{\T} := \{ {\Sigma}^{\T}_{(t)} : t \in \rman \}
\label{eq:time foliation}
\end{equation}
a {\it time foliation} of $\yman$. The time map $\T$ is required to
respect the orientation of $\yman$: if $t_1 < t_2$, the instant
${\Sigma}^{\T}_{(t_2)}$ has to lie in the future of
${\Sigma}^{\T}_{(t_1)}$ in $\yman$.

The space mapping ${\X}$ accompanying ${\T}$ is a mapping
\begin{equation}
{\X} \, : \, {\yman} \, \rightarrow \, {\xman} \, \, \, \, {\rm by}
\, \, \, \, y \in {\yman} \, \mapsto \, x = {\X}(y) \in {\xman}
\label{eq:intro of X}
\end{equation}
from $\yman$ to a three-dimensional manifold $\Sigma$, which is
assumed compact. The local coordinate representation of ${\X}$ is
${\X}^i$. The gradients ${\X}^i_{,\alpha}$ have to be spacelike;
i.e., for each $i$, we must have that $\iGab {\X}^i_{,\alpha}
{\X}^i_{,\beta} > 0$. Therefore, ${\X}$ has to depend on ${\bf
\Psi}$ as well. Each such mapping $\X$ associates a timelike
worldline $C^{\X}_{(x)}$ in $\yman$ with a point $x$ of $\xman$:
\begin{equation}
C^{\X}_{(x)} = \Big\{ y {\in} {\yman} : {\X}(y) = x {\in} {\xman}
\Big\} \, . \label{eq:timeline}
\end{equation}
This will be called {\it a reference worldline} and their collection
\begin{equation}
C^{\X} = \{ C^{\X}_{(x)} : x \in \xman \} \label{eq:reference frame}
\end{equation}
{\it a reference frame}.

The Cartesian product
\begin{eqnarray}
{\T} \times {\X} \; : \; {\yman} \, \, &\rightarrow& \, \, {\rman}
\times {\xman} \, \, \, \, {\rm by} \nonumber
\\
y \in {\yman} \, \, &\mapsto&  \, \, \big( t={\T}(y) \in {\rman} \,
,  \, x={ X}(y) \in {\xman}  \big) \label{eq:cartesian product}
\end{eqnarray}
associates the event $y \in \yman$ with the moment $t \in {\rman}$
and the point $x \in {\xman}$. Its inverse mapping
\begin{eqnarray}
{Y} \, : \, {\rman} \times {\xman} \, \, \rightarrow \, \, {\yman}
\, \, \, \, {\rm by}  \, \, \, \, \big( t \in {\rman} \, , \, x \in
{\xman} \big) \, \, \mapsto  \, \, y=Y(t,x) \in {\yman}
\label{eq:foliation}
\end{eqnarray}
will be called {\it the foliation mapping}. It may be viewed as a
one-parameter family $Y_{(t)}$, $t \in \rman$ of embeddings
\begin{equation}
{Y_{(t)}} \, : \, {\xman} \, \, \rightarrow \, \, {\yman} \, \, \,
\, {\rm by}  \, \, \, \, x \in {\xman} \, \, \mapsto  \, \,
{Y_{(t)}}(x) := Y(t,x) \in {\yman} \label{eq:one-param}
\end{equation}
of $\xman$ into $\yman$, whose images
${\Sigma}^{\T}_{(t)}={Y_{(t)}}(\xman)$ define the foliation
${\Sigma}^{\T}=\{ {\Sigma}^{\T}_{(t)} , \, t \in \rman \}$ of
$\yman$. It may also be viewed as a congruence $Y_{(x)}$, $x \in
\xman$ of curves
\begin{equation}
{Y_{(x)}} \, : \, {\rman} \, \, \rightarrow \, \, {\yman} \, \, \,
\, {\rm by}  \, \, \, \, t \in {\rman} \, \, \mapsto  \, \,
{Y_{(x)}}(t) := Y(t,x) \in {\yman} \; , \label{eq:congruence}
\end{equation}
whose images ${C}^{\X}_{(x)}={Y_{(x)}}(\rman)$ define the reference
frame $C^{\X}=\{ {C}^{\X}_{(x)} , x \in \xman \}$ in $\yman$. The
foliation mapping $Y$ locates the event $y \in \yman$ at which the
instant ${\Sigma}^{\T}_{(t)}=Y_{(t)}(\xman)$ intersects the
reference worldline ${C}^{\X}_{(x)}=Y_{(x)}(\rman)$.

\subsection{The transition from $\yman$ to $\xman \times \rman$}

Given an arbitrary configuration ${\bf \Psi}$ in ${\rm
Virt}{\yman}$, only a subset $\{ ({\T},{\X}) \}$ of mappings will
respect the light-cone structure induced on $\yman$ by the metric
${\G}$ in ${\bf \Psi}$. The requirements that ${\Sigma}^{\T}$ be
spacelike with respect to ${\G}$ and ${C}^{\X}$ be timelike with
respect to ${\G}$ imposes some restrictions on the induced fields
$\psi:={Y}^*{\bf \Psi}$ on $\xman \times \rman$. For example, the
pullback metric $g_{(t)}:={Y}_{(t)}^*{\G}$ induced on $\Sigma$ by
$\G$ and a given embedding $Y_{(t)}$ has to be positive definite for
all $t$. These restrictions on the fields $\psi$ on $\xman \times
\rman$ will be called {\it the compatibility conditions}, and the
corresponding mappings $\{(\T, \X)\}$ will be called {\it
compatible} with the configuration ${\bf \Psi}$ on $\yman$.

Following the presentation of the Dirac-ADM approach as given in Ref. \cite{time1}, the action is viewed as a functional of a reference configuration
${{\bf \Psi}}_o$ in ${\rm Virt}{\yman}$ and is pulled back from
$\yman$ to ${\Sigma} \times {\rm I}\!{\rm R}$ by a reference mapping
$Y_{o}$ whose associated time foliation ${\Sigma}^{{\T}_o}$ and
reference frame $C^{{\X}_o}$ are compatible with ${\G}_o \in {{\bf
\Psi}}_o$. When the action functional $S[{\bf \Psi}_o]$ on $\yman$
is pulled back to $\xman \times \rman$, the mapping $Y_o$ drops out
of the resulting functional, and the field equations are expressed
exclusively in terms of the induced configuration ${\psi}_o$ on
${\Sigma} \times {\rm I}\!{\rm R}$. This important feature enables
the theory to be expressed in the new context without the connecting
mappings $\T$, $\X$ and $Y$ taking any further part in its
formulation. In particular, the compatibility conditions for the
induced fields ${\psi}_o$ are turned into {\it definitions} for the
set ${\rm Virt}({\Sigma} \times {\rm I}\!{\rm R})$ of induced
virtual fields ${\psi}$ on ${\Sigma} \times {\rm I}\!{\rm R}$. For
example, the definition for the symmetric tensor $g_{(t)}$ on
${\Sigma}$ is that it has to be positive definite for all $t \in
\rman$. In this way, the action functional $S[{\psi}]$ is
subsequently varied independently of any link with $\yman$.

Assuming that four of the induced fields ${\psi}$ on ${\Sigma}
\times {\rm I}\!{\rm R}$ become Lagrange multipliers in the
canonical theory, the induced fields can be expressed in the
familiar form ${\psi}=(q,N,N^i)$. The lapse $N$ and the shift $N^i$
are respectively scalar and vector fields on $\Sigma$. They are both
scalar densities of weight one on ${\rm I}\!{\rm R}$. The dynamical
fields $q$, which include the induced metric $g$, may be chosen as
tensors on $\Sigma$ and as scalars on ${\rm I}\!{\rm R}$. Momenta
$p$ conjugate to $q$ are introduced. These are tensor densities of
weight one on $\Sigma$ and scalars on ${\rm I}\!{\rm R}$. The
Legendre transformation brings the Lagrangian action $S[q,N,N^i]$ on
$\Sigma \times {\rm I}\!{\rm R}$ into the canonical form
\begin{equation}
S[q,p;N,N^i] = \int dtd^3x (p\dot{q}-NH-N^iH_i) + \int dtd^3x
\dot{B} + \int dtd^3x {B^i}_{,i} \, . \label{split}
\end{equation}
The functionals $H(q,p)$ and $H_i(q,p)$ are the super-Hamiltonian
and super-momenta of the system. The boundary contributions $B$ and
$B^i$ depend on $N$, $N^i$, $q$ and $p$. Since $\xman$ has been
assumed compact, the spatial divergence vanishes.

The field equations on ${\Sigma} \times {\rm I}\!{\rm R}$ are
equivalent to Einstein's equations on $\yman$ in the following
sense: Given any solution ${\bf \Psi}$ of Einstein's equations and
any pair $({\T},{\X})$ that is compatible with $\G \in {\bf \Psi}$,
the induced fields $\psi$ satisfy the compatibility conditions and
the field equations on ${\Sigma} \times {\rm I}\!{\rm R}$.
Conversely, given any foliation mapping $Y$ and any configuration
$\psi$ in ${\rm Virt}({\rman} \times {\xman})$ that satisfies the
field equations on $\xman \times \rman$, the reconstructed
configuration ${\bf \Psi}$ satisfies Einstein's equations on $\yman$
and is such that the Cartesian product mapping ${\T} \times {\X}$
inverse to $Y$ is compatible with $\G \in {\bf \Psi}$.

The following results are also relevant, stated, for example, by
H\'{a}j\'{i}\v{c}ek and Kijowski \cite{HK}: Given a solution
manifold ${\yman}_{\bf \Psi}$ and any compatible mapping $\T \times
\X$, each hypersurface ${\xman}^{\T}_{(t_o)} \subset {\yman}_{\bf
\Psi}$, $t_o \in {\rman}$, is an admissible Cauchy surface; i.e., a
possible initial manifold for Einstein's equations. Let the initial
datum induced on $\xman$ by ${\bf \Psi}$ and the embedding
${Y}_{(t_o)} : \xman \rightarrow {\xman}^{\T}_{(t_o)}$ be $(q,p)$.
Then, this datum $(q,p)$ determines a unique class $\{ {\bf \Psi}
\}$ of ${\rm Diff}{\cal M}$-related solutions in ${\rm Sol}{\yman}$
in the following sense: For any choice of lapse $N$ and shift $N^i$,
and for any choice of foliation mapping $Y$, the maximal dynamical
development of $(q,p)$ reconstructs one of the solution manifolds in
the class $\{ {\bf \Psi} \}$.

\subsection{The reduced phase space $\Delta$ and the physical observables}

The phase space $\cal P$ may be defined as the cotangent bundle over
the set of the fields $q$ on $\Sigma$ equipped with a weak
symplectic form. Initial data appropriate for Einstein's equations
lie on the constraint surface ${\cal C} \subset {\cal P}$ determined
by $H(q,p)=0$ and $H^i(q,p)=0$. These data do not all determine
physically distinct states of the system. All points $(q,p)$ that
lie in an orbit of the Hamiltonian vector field generated by the
functional
\begin{equation}
H[q,p;N,N^i]= \int d^3x (NH-N^iH_i) \label{can generator}
\end{equation}
determine the same physical state. Specifically, for each choice of
the smearing functions $N$ and $N^i$, the functional (\ref{can
generator}) generates a one-parameter family of canonical
transformations of $\cal P$ that develops a given datum $(q,p)$ into
a curve of data in $\cal C$. The subset of ${\cal C}$ that can be
reached from this point $(q,p) \in {\cal C}$ by the action of
(\ref{can generator}) via all possible smearing functions defines
the orbit $o_{(q,p)}$ in $\cal C$. Since the datum $(q,p)$
determines a unique ${\rm Diff}{\cal M}$-class $\{ {\bf \Psi} \}$ of
solutions in ${\rm Sol}{\yman}$, it follows that each datum
$(q',p')$ in the orbit $o_{(q,p)}$ determines the same ${\rm
Diff}{\cal M}$-class $\{ {\bf \Psi} \}$ of solutions as $(q,p)$. It
is in this sense that it is can be claimed that the Hamiltonian
(\ref{can generator}) ``generates'' spacetime diffeomorphisms.

The quotient set $\Delta$,
\begin{equation}
\Delta := {\cal C}/o \, , \label{Dirac set}
\end{equation}
where $o$ denotes the orbits generated on ${\cal C}$ by (\ref{can
generator}), becomes the reduced phase space of the theory once the
symplectic form on $\cal P$ is pulled-back to $\Delta$. Each element
$\delta \in \Delta$ represents a ${\rm Diff}{\yman}$-class of
solutions in ${\rm Sol}{\yman}$. The correspondence between the sets
$\Gamma$ and $\Delta$ is therefore one-to-one, and the counting of
physically distinct states according to $\Gamma$ is recovered within
the canonical theory.

Regarding the physical observables, these are the Dirac observables.
Specifically, two functions $F_1(q,p)$ and $F_2(q,p)$ are considered
equivalent on ${\cal P}$ if their values are equal on the constraint
surface $\cal C$,
\begin{equation}
F_1(q,p){\mid}_{\cal C} = F_2(q,p){\mid}_{\cal C} \; .
\label{equivalent functions}
\end{equation}
Each such equivalence class of functions, represented by $F$,
defines a Dirac observable if, for all choices of $N$ and $N^i$, $F$
commutes on $\cal C$ with the Hamiltonian:
\begin{equation}
\{ F(q,p), H[q,p;N,N^i] \}{\mid}_{\cal C}=0 \, .
\label{D-observables}
\end{equation}
The Dirac observables $F$ on $\cal P$ can be projected via
(\ref{Dirac set}) to functionals on $\Delta$, as explained, for
example, by Fischer and Marsden \cite{FisMar}; they may also be
projected to functionals on $\Gamma$, as discussed by
H\'{a}j\'{i}\v{c}ek \cite{Hajicek}. In either case, these
observables remain invariant under the dynamical evolution generated
by the Hamiltonian. Since nothing physical actually evolves, this
evolution cannot be regarded as being measurable.

\section{Extending and reducing the action}

This section is concerned with the main concept behind the present approach.
Namely, it considers the procedure of incorporating non-dynamical variables into the variational principle of a ${\rm Diff}{\yman}$-invariant action. It investigates the effect of this procedure on the sets $\Gamma$ and $\Delta$ of the resulting action and the consequences this has for the emerging spacetime. In some respects, this topic is reminiscent of the issues raised by Kretschmann \cite{Kre}, Cartan \cite{Cartan} and Fock \cite{Fock}, when they discussed the extent to which Einstein's general theory obeys a relativity principle. Connections may also be recognised with the issues investigated by Anderson \cite{And}, Kucha\v{r} \cite{KucharKre} and more recently by Sorkin \cite{Sorkin}. However, no attempt will be made to relate the contents of this section to
one of these viewpoints.

\subsection{The method of extension}

In order to illustrate the extension of the Hilbert action without
the complications arising from the compatibility conditions between
$\T$, $\X$ and $\G$, it is best to put aside these conditions for
the moment, and incorporate them in the next section. Thus, assuming
that there is no relationship between $\T$, $\X$ and $\G$, one can
add these mappings to the Hilbert action,
\begin{equation}
S[\G] = {\int}_{{\yman}} d^4y \, \sqrt{-{\rm det}\G} \, R[\G] \, ,
\label{Hilbert's action}
\end{equation}
in a technically trivial way, without spoiling the general
covariance or the dynamical content of the field equations. There
are two stages in this procedure, which follows Ref. \cite{kk}:
First, the set ${\rm Virt}{\yman}={\rm Riem}{{\yman}}$ of virtual
fields of (\ref{Hilbert's action}) is extended by the set ${\rm
Maps}{\yman}$ of the time and space mappings, ${\T}:{\yman}
\rightarrow {\rman}$ and ${\X}:{\yman} \rightarrow {\xman}$, and
becomes the product set
\begin{equation}
{\rm Virt}{\yman} := {\rm Maps}{\yman} \times {\rm Riem}{{\yman}} \,
. \label{maps}
\end{equation}
Second, the action is viewed as a functional of all the variables
$\T$, $\X$ and $\G$,
\begin{equation}
S[\T,\X,\G] = {\int}_{{\yman}} d^4y \, \sqrt{-{\rm det}\G} \, R[\G]
\, , \label{Hilbert's action 2}
\end{equation}
without $\T$ and $\X$ actually appearing on the right hand side of
(\ref{Hilbert's action 2}).

In spite of the simplicity of this procedure, the distinction
between actions (\ref{Hilbert's action}) and (\ref{Hilbert's action
2}) is grounded on their sets $\Gamma$. In particular, the set ${\rm
Virt}{\yman}$ of virtual fields of (\ref{Hilbert's action 2}) is
larger than the corresponding set of (\ref{Hilbert's action}). In
addition, the variation of $\T$ and $\X$ in (\ref{Hilbert's action
2}) yields generally-covariant field equations---in fact,
identities---that leave $\T$ and $\X$ arbitrary. Therefore, virtual
fields and solutions for $\T$ and $\X$ are one and the same. If a
particular metric $\G$ solves Einstein's vacuum equations, i.e., if
$\G$ belongs to the set ${\rm Sol}^{\rm Hil}{\yman}$ of
(\ref{Hilbert's action}), then any pair ${\bf \Psi}=(\T,\X,\G)$,
where $(\T,\X) \in {\rm Maps}{\yman}$, belongs to the extended set
of solutions ${\rm Maps}{\yman} \times {\rm Sol}^{\rm Hil}{\yman}$
of (\ref{Hilbert's action 2}). As a result, the set $\Gamma$ of
${\rm Diff}{\yman}$-classes of solutions of (\ref{Hilbert's action
2}) is larger than that of (\ref{Hilbert's action}).

Furthermore, the fact that the variables $\T$ and $\X$ are absent
from the right hand side of (\ref{Hilbert's action 2}) cannot be
used as an argument for rejecting (\ref{Hilbert's action 2}) as a
genuine action. The reason is that a different choice of variables
in the set ${\rm Virt}{\yman}$ of (\ref{Hilbert's action 2}) can
make the right hand side of (\ref{Hilbert's action 2}) depend
explicitly on all these variables. For example, one may choose the
ADM chart on ${\rm Virt}{\yman}$ consisting of the variables ${\T}$,
${\X}$, ${\bf N}$, ${\bf N}^i$ and ${\bf g}_{ij}$, where the
boldface lapse ${\bf N}$, shift ${\bf N}^i$ and spatial metric ${\bf
g}_{ij}$ are viewed as scalar functions on $\yman$. In this case,
action (\ref{Hilbert's action 2}) becomes
\begin{equation}
S[{\T},{\X},{\bf N},{\bf N}^i,{\bf g}_{ij}] = {\int}_{{\cal M}} d^4y
\, |{{\partial ({\T},{\X})}\over{\partial y^{\alpha}}}| \, {\bf N}
\sqrt{{\rm det}{\bf g}} \Big( {\bf k}_{ij} {\bf k}^{ij} - ( {\bf
k}^i_i)^2 + R[{\bf g}] \Big) \, . \label{Hilbert's action 3}
\end{equation}
The boldface extrinsic curvature ${\bf k}_{ij}$ and the spatial
curvature scalar $R[{\bf g}]$ are also viewed as scalar functions on
$\yman$. The Jacobian $|{{\partial ({\T},{\X})}\over{\partial
y^{\alpha}}}|$ in (\ref{Hilbert's action 3}) depends on the mappings
$\T$ and $\X$ and, therefore, all the elements of the set ${\rm
Virt}{\yman}$ are present in this version of the action. In particular, the variation of (\ref{Hilbert's action 3}) produces equations that are all generally-covariant and non-trivial. Thus, despite appearances, actions (\ref{Hilbert's action}) and (\ref{Hilbert's action 2}) are distinct in agreement with their distinct sets $\Gamma$, while actions (\ref{Hilbert's action 2}) and (\ref{Hilbert's action 3}) are just equivalent.

If now physical significance is attributed to the set $\Gamma$, the process of extension of the Hilbert action by non-dynamical variables has observable consequences. And, dragging this argument a little further, the principle of general covariance cannot, by itself, stop such a proliferation of observables. For example, a variety of spacetime fields may be added to the set ${\rm Virt}{\yman}$ of (\ref{Hilbert's action}), without any of them appearing in the action. All these fields will remain undetermined by the variational principle, and this will produce yet another set $\Gamma$ of observables, while leaving the general covariance of the formalism intact. Under these circumstances, the immediate need arises for establishing a criterion that can decide on the appropriateness of the selected variables. Given that the principle of general covariance is not such a criterion, one has to resort to the interpretation of the chosen variables, and require that this interpretation be consistent with observation; i.e., that any predicted observables be actually observed. Seen in this light, the formalism based on (\ref{Hilbert's action 2}) is clearly inappropriate, because the mappings $({\T},{\X})$ are entirely independent of $\G$, while reference systems are causal.

\subsection{Reduction via Dirac's requirement}

Before modifying (\ref{Hilbert's action 2}) by introducing
compatibility conditions between $\T$, $\X$ and $\G$, it is worth
considering the reverse procedure; namely, the reduction from
(\ref{Hilbert's action 2}) to the original action (\ref{Hilbert's
action}). Although these formalisms are distinct according to their
sets $\Gamma$, it is also true that they look similar. This
similarity is established by the Bergmann-Dirac analysis of the
first-class constraints. More precisely, while the differences
between (\ref{Hilbert's action}) and (\ref{Hilbert's action 2}) are
expressed by the set $\Gamma$ of ${\rm Diff}{\yman}$-classes of
solutions, their similarities are expressed by the set $\Delta$ of
first-class orbits.

Actions (\ref{Hilbert's action}) and (\ref{Hilbert's action 2})
yield sets $\Delta$ that are isomorphic. In the case of
(\ref{Hilbert's action}), the correspondence between the sets
$\Gamma$ and $\Delta$ is bijective, allowing the Hamiltonian and
momentum constraints to be associated with ${\rm Diff}{\cal M}$. In
the case of (\ref{Hilbert's action 2}), any canonical analysis that
incorporates the non-dynamical variables $\T$ and $\X$ in the phase
space will yield eight first-class constraints, four too many to be
associated with ${\rm Diff}{\cal M}$. The remaining four first-class
constraints reflect the fact that the mappings $\T$ and $\X$ are
left undetermined by the variational principle. These extra
constraints imply that the set $\Delta$ of (\ref{Hilbert's action
2}) is a subset of its set ${\Gamma}$, and isomorphic to the set
$\Delta$ of (\ref{Hilbert's action}).

Now, if the point of view is followed that the observable aspects of
a generally-covariant theory are associated solely with its set
${\Delta}$, then some of the physical premises of (\ref{Hilbert's
action 2}), according to its set $\Gamma$, have to be invalidated.
This reduction from the physical premises of (\ref{Hilbert's action
2}) to those of (\ref{Hilbert's action}) can be achieved by
postulating, following Bergmann and Dirac, that all eight
first-class constraints in the phase space of (\ref{Hilbert's action
2}) generate gauge transformations. This is sometimes referred to as
Dirac's conjecture---analysed in detail in Refs. \cite{Diracqm},
\cite{HT} and \cite{Pons}---although it may be regarded as a
physical requirement.

Imposing this requirement on the formalism defined by
(\ref{Hilbert's action 2}) has the following consequences: If all
first-class constraints generate gauge transformations, then taking
the quotient of the set of solutions of (\ref{Hilbert's action 2})
by the orbits of ${\rm Diff}{\cal M}$ is not sufficient to eliminate
all gauge freedom. It is necessary to take the quotient of the
resulting set ${\Gamma}$ by the orbits in ${\Gamma}$ generated by
arbitrary changes of the fields ${\T}$ and ${\X}$. These changes are
symmetries of the field equations derived from (\ref{Hilbert's
action 2}). The fact that these symmetries are physically
unimportant can be deduced from the analysis of the first-class
constraints and the transformations these generate in the phase
space. The quotient of ${\Gamma}$ by the orbits generated by these
symmetries yields a set that is isomorphic to the set ${\Delta}$.

\section{The extended action}

The ability to reduce the physical content of the extended formalism
to that of the conventional formalism by making use of
Dirac's requirement means that it is not necessary to commit to a
particular interpretation. Both options for interpreting canonical
vacuum gravity are incorporated in the extended action. Thus, in
this section, the compatibility conditions between $\T$, $\X$ and
$\G$ are added to action (\ref{Hilbert's action 2}), and the
covariant canonical formalism is derived. By necessity, the
following four sections are technical. However, the conceptual
characteristics of the formalism are discussed again in the last section.

\subsection{The Lagrangian on $\yman$}

Given coordinates $y^{\alpha}$ on $\yman$, $x^i$ on $\xman$, and $t$
on $\rman$, the metric field $\Gab$ is pulled back to $\xman \times
\rman$ by the foliation mapping $Y$ to yield the induced metric
$g_{ij}$, the shift vector $N^i$ and the lapse function $N$
\begin{eqnarray}
\Gab(Y) {Y}^{\alpha}_{,i} {Y}^{\beta}_{,j} &:=& g_{ij} \, ,
\label{GXX}
\\
\Gab(Y) {Y}^{\alpha}_{,i} {Y}^{\beta}_{,t} &:=& g_{ij} N^j  \, ,
\label{GXT}
\\
\Gab(Y) {Y}^{\alpha}_{,t} {Y}^{\beta}_{,t} &:=& g_{ij} N^i N^j -N^2
\, . \label{GTT}
\end{eqnarray}
Alternatively, $\G$ may be expanded in the $\G$-independent co-basis
constructed from the mappings $\T$ and $\X$, according to
\begin{equation}
\Gab = \big( {\bf g}_{ij}{\bf N}^i {\bf N}^j - {\bf N}^2 \big)
{\T}_{, \alpha} {\T}_{,\beta} + {\bf g}_{ij} {\bf N}^j  \big(
{\T}_{, \alpha}{\X}^i_{,\beta} + {\X}^i_{, \alpha} {\T}_{, \beta}
\big) + \big( {\bf g}_{ij} {\X}^i_{,\alpha} {\X}^j_{,\beta} \big)
\, , \label{G}
\end{equation}
where the boldface lapse, shift and spatial-metric are viewed as
fields on $\yman$:
\begin{eqnarray}
{\bf g}_{ij} := g_{ij}\big({\T},{\X}\big) \, , \label{bfg}
\\
{\bf N} :=N\big({\T},{\X}\big) \, , \label{bfN}
\\
{\bf N}^i :=N^i\big({\T},{\X}\big) \, . \label{bfNi}
\end{eqnarray}

Equation (\ref{G}), which provides the link between the spacetime
fields and the fields on ${\rman} \times {\xman}$, is adjoined to
the extended action (\ref{Hilbert's action 2}) by Lagrange
multipliers. The independent variables in the resulting action
\begin{equation} S[{\G},{\T},{\X},N,N^i,g,{\bf M}] = {\int} d^4X \,
\sqrt{-{\rm det}\G} \, \big( R[\G] + {\bf M}^{\alpha \beta} \,
{C}_{\alpha \beta} \big) \label{maxsym}
\end{equation}
are chosen to be the metric field $\G$, the mappings $\T$ and $\X$,
the induced fields $N$, $N^i$ and $g$ on $\xman \times \rman$ and
finally the symmetric tensor density multipliers ${\bf M}$. The
tensor field
\begin{equation}
{C}_{\alpha \beta} := \G_{\alpha \beta} - \big( {\bf g}_{ij} {\bf
N}^i {\bf N}^j - {\bf N}^2 \big) {\T}_{, \alpha} {\T}_{, \beta} -
{\bf g}_{ij} {\bf N}^j \big( {\T}_{, \alpha} {\X}^i_{, \beta} +
{\X}^i_{, \alpha} {\T}_{, \beta} \big) - {\bf g}_{ij} {\X}^i_{,
\alpha} {\X}^j_{, \beta} \, , \label{C}
\end{equation}
is just a re-arrangement of (\ref{G}), where the boldface lapse,
shift and induced metric are viewed as functionals of the
independent variables $N$, $N^i$, $g_{ij}$, $\T$ and $\X$ via
(\ref{bfg})-(\ref{bfNi}). The space of virtual fields of
(\ref{maxsym}) incorporates the appropriate definitions for the
induced fields $(g, N, N^i)$ on $\xman \times \rman$ that ensure
compatibility between $\T$, $\X$ and $\G$ at the level of the
solutions. On the other hand, the mappings $\T$ and $\X$ are
unrestricted prior to variation. The manifest invariance of the
action (\ref{maxsym}) under the diffeomorphisms of all manifolds
$\yman$, $\rman$ and $\xman$ should be noted.

The dynamical content of the equations derived from (\ref{maxsym})
is equivalent to that of Einstein's equations for vacuum gravity. In
particular, the variation of (\ref{maxsym}) with respect to the
multipliers ${\bf M}$ yields the constraint (\ref{G}), relating the
spacetime fields with the fields on ${\rman} \times {\xman}$. Since
the compatibility conditions have been incorporated in the set ${\rm
Virt}{\yman}$ of (\ref{maxsym}), constraint (\ref{G}) ensures that
the mappings $\T$ and $\X$ are compatible with $\G$ at the level of
the solutions. The variation of (\ref{maxsym}) with respect to $N$,
$N^i$ and $g$ implies that all projections of the multipliers ${\bf
M}$, and hence ${\bf M}$ themselves, vanish:
\begin{eqnarray}
{\bf M}^{\alpha \beta}{\T}_{, \alpha}{\T}_{, \beta} &=& 0 \, ,
\label{MTT}
\\
{\bf M}^{\alpha \beta} \big( {\T}_{, \alpha}{\X}^i_{, \beta} +
{\X}^i_{, \alpha}{\T}_{, \beta} \big) &=& 0 \, , \label{MTX}
\\
{\bf M}^{\alpha \beta}{\X}^i_{, \alpha}{\X}^j_{, \beta} &=& 0 \, .
\label{MXX}
\end{eqnarray}
Regarding the variation of $\G$, the contributions arising from the
second term in (\ref{maxsym}) are proportional to the vanishing
multipliers ${\bf M}$. Therefore, the vacuum Einstein equations are
preserved:
\begin{equation}
R^{\alpha \beta}[{\G}] - {1\over2} R[{\G}]{\G}^{\alpha \beta} = 0 \,
. \label{Einstein's}
\end{equation}
Finally, the variations of the time mapping $\T$ and the space
mapping $\X$ yield equations that are valid by means of the
remaining field equations, and hence provide no additional
information.

\subsection{The Lagrangian on ${\rman} \times {\xman}$}

Following \cite{kk}, the transition from $\yman$ to $\xman \times
\rman$ is viewed as a change of ``coordinate chart'' on the set of
virtual fields of (\ref{maxsym}): The pair of mappings $(\T,\X)$ is
replaced by its inverse mapping $Y$, the multiplier $\bf M$ is
replaced by its components $(M,M^i,M^{ij})$ in the $\G$-independent
basis on $\yman$ constructed from $Y$,
\begin{equation}
M Y^{\alpha}_{,t} Y^{\beta}_{,t} + {1\over2} M^i \big(
Y^{\alpha}_{,t} Y^{\beta}_{,i} + Y^{\alpha}_{,i} Y^{\beta}_{,t}
\Big) + M^{ij} Y^{\alpha}_{,i} Y^{\beta}_{,j}  := {\bf M}^{\alpha
\beta}(Y) \, ,
\end{equation}
and the action (\ref{maxsym}) is pulled back to an equivalent
Lagrangian action on $\xman \times \rman$:
\begin{equation}
S[{\G},Y,N,N^i,g,M,M^i,M^{ij}] = {\int} d^3xdt N \sqrt{{\rm det}g}
\Big( {\kappa}_{ij} {\kappa}^{ij} - ( {\kappa}^i_i)^2 + R[g] + M C +
M^i C_i +M^{ij} C_{ij} \Big) \, . \label{maxsym2}
\end{equation}
The constraints
\begin{eqnarray}
C &:=& {\Gab}(Y) Y^{\alpha}_{,t} Y^{\beta}_{,t} - g_{ij} N^i N^j +
N^2  \, , \label{constraint1}
\\
C_i &:=& {\Gab}(Y) Y^{\alpha}_{,t} Y^{\beta}_{,i} - g_{ij} N^j \, ,
\label{constraint2}
\\
C_{ij} &:=& {\Gab}(Y) Y^{\alpha}_{,i} Y^{\beta}_{,j} - g_{ij}
\label{constraint3}
\end{eqnarray}
reproduce definitions (\ref{GXX})-(\ref{GTT}), $\kappa$ is the
extrinsic curvature
\begin{equation}
{\kappa}_{ij} := {1\over{2N}} \big( -{\dot{g}}_{ij} + N_{i;j}
+N_{j;i} \big) \, , \label{extrinsic}
\end{equation}
and $R[g]$ is the curvature scalar. The constraints
(\ref{constraint1})-(\ref{constraint3}) have been already used to
eliminate all other occurrences of $\G$ in the action functional; a
procedure that is always permitted.

Now, in general, a set of variables can be eliminated completely
from the formalism as long as the field equations obtained by
varying these variables in the action functional determine the
varied variables uniquely. The relevant information is preserved by
turning these equations into definitions for the eliminated
variables. In the case of (\ref{maxsym2}), the simultaneous
variation of the multipliers $(M, M^i , M^{ij})$ and the metric
field $\G$ implies that these multipliers vanish,
\begin{equation}
M=0 \, , \; \; \; \; \; \; \; \; M^i =0 \, , \; \; \; \; \; \; \; \;
M^i =0 \, , \label{MMM}
\end{equation}
and determines $\G$ uniquely in terms of the remaining variables
according to (\ref{GXX})-(\ref{GTT}). A key property of this
procedure is that, although it reduces the size of the set of
virtual fields, it preserves the subset of solutions and
consequently the set $\Gamma$ of ${\rm Diff}{\yman}$-classes of
solutions. This follows from the uniqueness of its outcome; i.e.,
from the fact that the eliminated variables are expressed uniquely
in terms of the variables that remain in the theory. In the
particular case of (\ref{maxsym2}), after the redundant variables
$M$, $M^i$, $M^{ij}$ and $\G$ are eliminated, the mapping $Y$ drops
out of the right hand side of the simplified action as well:
\begin{equation}
S[Y,N,N^i,g] = {\int} d^3xdt N \sqrt{{\rm det}g} \Big( {\kappa}_{ij}
{\kappa}^{ij} - ( {\kappa}^i_i)^2 + R[g] \Big) \, . \label{maxsym3}
\end{equation}
However, it should not be forgotten that this mapping is still
present in the set of virtual fields of (\ref{maxsym3}), reflecting
the fact that the set $\Gamma$ of (\ref{maxsym3}) is larger than the
corresponding set of the Hilbert action. As a confirmation of the
consistency of this reasoning, $Y$ cannot be eliminated by means of
its own variation in (\ref{maxsym3}), since this variation yields an
identity which is unable to determine $Y$ uniquely.

\subsection{The covariant canonical action}

Momenta conjugate to both $g$ and $Y$ need to be introduced. The
Legendre transformation followed by the Dirac analysis of the
primary constraints brings (\ref{maxsym3}) into the first-class
canonical form
\begin{equation}
S[Y,P,q,p;N,N^i,{\Lambda}] = \int dtd^3x \big( P_{\alpha}
{Y}^{\alpha}_{,t} + p^{ij} q_{ij,t} - NH - N^iH_i -
{\Lambda}^{\alpha}P_{\alpha} \big) \, , \label{maxsym4}
\end{equation}
where ${\Lambda}^{\alpha}$, $N$ and $N^i$ are Lagrange multipliers.
In contrast to the mappings $\T$ and $\X$, which satisfy the
compatibility constraint (\ref{C}), the mapping $Y$ remains entirely
arbitrary after the variation of (\ref{maxsym4}). This is because,
in this version of the extended action, the compatibility conditions
are fully integrated into the set of virtual fields on ${\rman}
\times {\xman}$ and, hence, are transferred directly to the level of
the solutions. In particular, since $g \in {\rm Virt}{\yman}$ is
positive definite, and $N\in {\rm Virt}{\yman}$ is positive, any
solution $(g,p)$ of (\ref{maxsym4}) combined with any foliation
mapping $Y$ reconstruct a spacetime metric $\G$ which is, by
construction, compatible with the mappings $\T$ and $\X$ inverse to
$Y$.

Thus, the variation of (\ref{maxsym4}) yields the standard equations
of the Dirac-ADM action,
\begin{eqnarray}
g_{ij,t} &=& \{g_{ij} , \int d^3x (NH + N^mH_m +
{\Lambda}^{\beta}P_{\beta}) \} = \{g_{ij} , \int d^3x (NH + N^mH_m)
\}     \, , \label{Ham equ1}
\\
{p^{ij}}_{,t} &=& \{p^{ij} , \int d^3x (NH + N^mH_m +
{\Lambda}^{\beta}P_{\beta}) \} = \{p^{ij} , \int d^3x (NH + N^mH_m)
\}    \, , \label{Ham equ2}
\\
H &=& 0 \, , \label{Ham con}
\\
H_i &=&  0 \, , \label{Mom con}
\end{eqnarray}
supplemented by the expected equations for the foliation variables:
\begin{eqnarray}
Y^{\alpha}_{,t} &=& \{ Y^{\alpha} , \int d^3x (NH + N^mH_m +
{\Lambda}^{\beta}P_{\beta}) \} = {\Lambda}^{\alpha} \, , \label{Ham
equ1B}
\\
P_{\alpha,t} &=& \{ P_{\alpha} , \int d^3x (NH + N^mH_m +
{\Lambda}^{\beta}P_{\beta}) \} = 0   \, , \label{Ham equ2B}
\\
P_{\alpha} &=& 0 \, . \label{conB}
\end{eqnarray}

\section{Equivalence classes of solutions and the sets $\Gamma$ and $\Delta$}

This section investigates the set of solutions, the set of ${\rm
Diff}{\yman}$-classes of solutions, and the set of first-class
orbits of the extended action. The many-to-one relationship that
exists between the last two sets forms the backbone of the covariant
canonical formalism.

\subsection{The set of solutions}

Let us first summarise the content of the extended formalism. The
spacetime action (\ref{maxsym}) depends on the metric ${\G}$, the
mappings ${\T}$ and ${\X}$, the induced fields $N$, $N^i$ and $g$ on
${\rman} \times {\xman}$ and the multipliers ${\bf M}$. The set
${\rm Virt}{\yman}$ of all kinematically admissible configurations
of the theory includes the appropriate restrictions for the fields
$N$, $N^i$ and $g$ on ${\rman} \times {\xman}$ in order to ensure
compatibility between $\T$, $\X$ and $\G$ at the level of the
solutions. The field equations derived from (\ref{maxsym}) imply
that the multipliers $\bf M$ vanish, the metric $\G$ satisfies the
Einstein vacuum equations, and the remaining fields $\T$, $\X$, $N$,
$N^i$ and $g$ are related to the solution $\G$ through equation
(\ref{G}). Therefore, each solution $s \in {\rm Sol}{\yman}$ can be
equivalently described by the set of fields
\begin{equation}
s = ({\G}, {\T}, {\X}) \, , \label{solution of m}
\end{equation}
provided that $\G$ satisfies the Einstein vacuum equations, and $\T$
and $\X$ are compatible with $\G$ in the sense that their gradients
are respectively timelike and spacelike. Within the framework of the
canonical action (\ref{maxsym4}), each solution $s \in {\rm
Sol}({\rman} \times {\xman})$ can be described by the set
\begin{equation}
s = (g,p,Y) \, , \label{solution of can}
\end{equation}
provided that $g$ and $p$ and $Y$ satisfy the field equations of
(\ref{maxsym4}). The connection between the spacetime and the
canonical representations of solutions is the following: Given a
solution $(g,p,Y) \in {\rm Sol}({\rman} \times {\xman})$ of
(\ref{maxsym4}), the lapse $N$ and the shift $N^i$ can be in
principle calculated from $g$ and $p$, determining a solution
\begin{equation}
s = (g,N,N^i,Y) \label{solution of rt}
\end{equation}
of the field equations of the Lagrangian action (\ref{maxsym3}).
Then, the solution $({\G}, {\T}, {\X}) \in {\rm Sol}{\yman}$ of the
spacetime action (\ref{maxsym}), constructed from $(g, N, N^i, Y)$,
is such that the mappings ${\T}$ and ${\X}$, inverse to $Y$, are
compatible with ${\G}$.

\subsection{The set $\Gamma$}

Regardless of representation, an equivalence relationship exists
between the solutions of any version of the extended action. In the
context of the spacetime action (\ref{maxsym}), two solutions
$({\G}_{1}, {\T}_{1}, {\X}_{1})$ and $({\G}_{2}, {\T}_{2},
{\X}_{2})$ are equivalent,
\begin{equation}
({\G}_{1}, {\T}_{1}, {\X}_{1}) \sim ({\G}_{2}, {\T}_{2}, {\X}_{2})
\, , \label{equivalence 1}
\end{equation}
if there exists a diffeomorphism $D: {\yman} \rightarrow {\yman}$
such that
\begin{equation}
{\G}_2=D_{*}{\G}_1 \, ,  \, {\T}_2=D_{*}{\T}_1 \, , \,
{\X}_2=D_{*}{\X}_1 \, , \label{diffm-invariance 2}
\end{equation}
where $D_{*}$ is the push-forward mapping. Each equivalence class of
solutions determines a unique element $\gamma \in \Gamma$, which can
be denoted by \begin{equation} \gamma=\{ ({\G},{\T},{\X}) \} \, .
\label{gamma 1}
\end{equation}
In the context of the canonical action (\ref{maxsym4}), the
classification of the elements of $\Gamma$ looks actually simpler:
Two solutions $(g_{1},p_{1},Y_{1})$ and $(g_{2},p_{2},Y_{2})$ of
(\ref{maxsym4}) are equivalent,
\begin{equation}
(g_{1},p_{1},Y_{1}) \sim (g_{2},p_{2},Y_{2}) \, , \label{equivalence
2}
\end{equation}
if they have the same components $g$ and $p$,
\begin{equation}
g_{1}=g_{2} \, , \,  p_{1}=p_{2} \, , \label{equivalence 3}
\end{equation}
and differ only in their component $Y$. This fact follows from the
covariance of the fields $\G$, $\T$ and $\X$ under spacetime
diffeomorphisms, which leads to the invariance of the fields $g$ and
$p$. Hence, each element $\gamma \in \Gamma$ can be denoted by
\begin{equation}
\gamma=(g,p) \, , \label{gamma 2}
\end{equation}
or, equivalently, by
\begin{equation}
\gamma=(g,N,N^i) \, . \label{gamma 3}
\end{equation}
It should be noticed that the set $\Gamma$ of the extended action is
isomorphic to the set ${\rm Sol}^{\rm ADM}{({\rman} \times
{\xman})}$ of solutions of the Dirac-ADM action.

\subsection{The set $\Delta$}

The set $\Delta$ of the extended action coincides with the
corresponding set of the Dirac-ADM action due to the presence of the
additional first-class constraints in (\ref{maxsym4}). The
equivalence relation that connects the sets $\Gamma$ and $\Delta$ of
(\ref{maxsym}) is, as expected, the following: Two elements
${\gamma}_{1}=\{ ({\T}_{1},{\X}_{1},{\G}_{1}) \}$ and
${\gamma}_{2}=\{ ({\T}_{2},{\X}_{2},{\G}_{2})\}$ of the set $\Gamma$
are equivalent,
\begin{equation}
{\gamma}_{1} \sim {\gamma}_{2} \, , \label{equivalence 20}
\end{equation}
if the metric classes contained in ${\gamma}_{1}$ and ${\gamma}_{2}$
are the same,
\begin{equation}
\{{\G}_{1}\} = \{{\G}_{2}\} \, . \label{equivalence 30}
\end{equation}
In other words, in order to obtain the set $\Delta$ from the set
$\Gamma$ of the extended action one has to treat the mappings $\T$
and $\X$ as {\it unimportant} elements of the theory.

In the canonical representation, the same applies to the foliation
mapping $Y$. Specifically, two elements ${\gamma}_{1}=(g_1,p_1)$ and
${\gamma}_{1}=(g_1,p_1)$ of the set $\Gamma$ are considered
equivalent,
\begin{equation}
(g_1,p_1)=(g_2,p_2) \, , \label{equivalence 40}
\end{equation}
if the classes $\{ {\G}_1 \}$ and $\{ {\G}_1 \}$, constructed from
$(g_1,p_1)$ and $(g_2,p_2)$ via {\it arbitrary} choices of the
mapping $Y$, are the same. Thus, the relegation of the fields $\T$,
$\X$ and $Y$ of the extended action to secondary elements of the
theory is equivalent, in meaning, to the standard treatment of the
foliation as an {\it external} element of the Dirac-ADM theory.

Summarising the findings of this section, the set $\Delta$ of the
extended action is isomorphic to the set ${\Delta}^{\rm ADM}$ of the
Dirac-ADM action and the set ${\Gamma}^{\rm Hil}$ of the Hilbert
action. However, the set $\Gamma$ of the extended action is
enlarged, because each element $\delta \in \Delta$ represents a
whole class $\{ \gamma \}$ of elements in $\Gamma$, making the
correspondence between the sets $\Gamma$ and $\Delta$ many-to-one.

\section{Symmetries}

A generalised symmetry of the field equations is an infinitesimal
transformation of the set ${\rm Virt}{\yman}$ that preserves its
subset ${\rm Sol}{\yman}$ of solutions. In general, the generator
${\Delta}{{\bf \Psi}}[{\bf \Psi}]$ of this transformation is a
functional of all the field variables ${\bf \Psi}$. This generator
satisfies the linearised form of the field equations provided that
these equations hold. Such symmetries map solutions in ${\rm
Sol}{\yman}$ to other `neighbouring' solutions. All local
generalised symmetries of vacuum general relativity have been found
by Anderson and Torre \cite{AndTor}. These consists of generalised
spacetime diffeomorphisms and the constant scaling of the spacetime
metric. Of course, the extension of the Hilbert action by the
mappings $\T$ and $\X$ implies additional symmetries, the most
important of which will be investigated in this section.

Overall, the symmetries that will be examined are split into two
categories. The first category consists of symmetries induced by the
diffeomorphisms of the manifolds $\yman$, $\xman$ and $\rman$. These
are generated by vector fields $V^{\alpha}$, $v^i$ and $w$, that are
regarded, respectively, as elements of the Lie algebras ${\rm L}{\rm
Diff}{\yman}$, ${\rm L}{\rm Diff}{\xman}$ and ${\rm L}{\rm
Diff}{\rman}$. Each of these groups of diffeomorphisms moves the
elements of its own manifold and keeps the elements of the other
manifolds fixed. Consequently, the transformations of the connecting
mappings $\T$, $\X$ and $Y$ are especially important because they
determine how the fields on $\yman$ transform under ${\rm
Diff}{\xman}$ and ${\rm Diff}{\rman}$, and how the fields on ${\xman
\times \rman}$ transform under ${\rm Diff}{\yman}$.

The second category of symmetries consists of generalised changes
$\delta{\T}[{\bf \Psi}]$ and $\delta{\X}[{\bf \Psi}]$, ${\bf
\Psi}=({\G},{\T},{\X})$, of the time and space mappings, where the
events $y \in \yman$ and the spacetime metric $\G$ are kept fixed,
as well as generalised changes $\delta{Y}[{g},{p},{Y}]$ of the
foliation mapping, where the moments $t \in {\rman}$, the points $x
\in {\xman}$ and the fields $g$ and $p$ are kept fixed. There are
certain similarities between these two categories of symmetries
which will be spelled out towards the end of this section. For
simplicity, only the main variables will be considered; namely the
spacetime fields ${\G}_{{\alpha}{\beta}}$, ${\T}$ and ${\X}$, the
induced fields $g$, $N$, $N^i$, and $Y$, the momenta $p$ and $P$,
and the canonical multipliers ${\Lambda}$. The remaining multipliers
$M^{\alpha\beta}$, $M^{ij}$, $M^i$ and $M$ are not present in the
covariant canonical action (\ref{maxsym4}), so they will be left out
of this discussion.

\subsection{Infinitesimal symmetries induced by ${\rm Diff}{\yman}$}

These symmetries are generated by vector fields $V^{\alpha}$ on
$\yman$, and are defined by
\begin{eqnarray}
\delta y^{\alpha}  &:=&  V^{\alpha}  \, , \label{ychanges 01}
\\
\delta x^i &:=& 0         \, , \label{ychanges 02}
\\
\delta t &:=& 0        \, . \label{ychanges 03}
\end{eqnarray}
The spacetime fields respond to these changes via their Lie
derivative ${\cal L}_{V}$:
\begin{eqnarray}
\delta {\G}_{\alpha \beta} &=& - {\cal L}_{V}{\G}_{\alpha \beta} =
-{\G}_{\alpha \beta,\gamma}V^{\gamma} -{\G}_{\alpha \gamma}
V^{\gamma}_{,\beta} -{\G}_{\gamma \beta} V^{\gamma}_{,\alpha} \, ,
\label{ychanges 1}
\\
\delta {\T} &=& - {\cal L}_{V}{\T} = -{\T}_{,\gamma}V^{\gamma} \, ,
\label{ychanges 2}
\\
\delta {\X}^i &=& - {\cal L}_{V}{\X}^i = -{\X}^i_{,\gamma}V^{\gamma}
\, . \label{ychanges 3}
\end{eqnarray}
The response of the connecting mapping $Y$ is determined by the
condition $Y({\T},{\X})={\rm Id}_{\yman}$, where ${\rm Id}_{\yman}$
is the identity transformation of $\yman$. Infinitesimally, this
relation becomes
\begin{equation}
{\delta}Y^{\alpha} = -Y^{\alpha}_{,t} \delta{\T}(Y) -
Y^{\alpha}_{,i} \delta{\X}^i(Y) \, . \label{YXT}
\end{equation}
Using (\ref{ychanges 2})-(\ref{YXT}) and the definitions
(\ref{GXX})-(\ref{GTT}) of the fields $g$, $N$ and $N^i$ in terms of
the spacetime variables, the following changes are determined:
\begin{eqnarray}
\delta Y^{\alpha} &=& V^{\alpha}(Y) \, , \label{ychanges 4}
\\
\delta g_{ij} &=& 0 \, , \label{ychanges 5}
\\
\delta N &=& 0 \, , \label{ychanges 6}
\\
\delta N^{i} &=& 0 \, . \label{ychanges 7}
\end{eqnarray}
In addition, the field equations derived from the canonical theory
(\ref{maxsym4}) and the relations (\ref{ychanges 5})-(\ref{ychanges
7}) imply that the momenta $p$ remain unchanged:
\begin{equation}
\delta p^{ij} = 0 \, . \label{ychanges 8}
\end{equation}
Finally, regarding the embedding momenta $P$ and the Lagrange
multipliers $\Lambda$ in (\ref{maxsym4}), one has to rely again on
the field equations. The following changes then arise:
\begin{eqnarray}
\delta P_{\alpha} &=& 0 \, , \label{ychanges 11}
\\
\delta {\Lambda}^{\alpha} &=& {\Lambda}^{\beta}
V^{\alpha}_{,\beta}(Y) \, . \label{ychanges 12}
\end{eqnarray}

The geometric meaning of these changes, with the emphasis placed on
the connecting mappings, was originally discussed in Ref. \cite{kk}:
While spacetime diffeomorphisms do not move the moments $t$ in
$\rman$, they change the time map $\T$ and hence the foliation
${\Sigma}^{\T}$ in $\yman$. They send the instant
${\Sigma}^{\T}_{(t)}$ of the original time foliation ${\Sigma}^{\T}$
onto the instant ${\Sigma}^{{\T}+{\delta{\T}}}_{(t)}$ of a different
time foliation ${\Sigma}^{{\T}+{\delta}{\T}}$: In general, the
instant ${\Sigma}^{{\T}+{\delta}{\T}}_{(t)}$ is not only different
from the instant ${\Sigma}^{\T}_{(t)}$, but also from all the other
instants ${\Sigma}^{\T}_{(t')}$, $t' \in {\rman}$, of the original
time foliation ${\Sigma}^{\T}$. Similarly, while spacetime
diffeomorphisms do not move the points $x$ in $\xman$, they change
the map $\X$ and hence the reference frame $C^{\X}$ in $\yman$. They
send the reference worldline $C^{X}_{(x)}$ of the original reference
frame $C^{\X}$ onto the reference worldline
${C}^{{\X}+{\delta}{\X}}_{(x)}$ of a different reference frame
$C^{{\X}+\delta{\X}}$: In general, the reference worldline
${C}^{{\X}+\delta{\X}}_{(x)}$ is not only different from the
reference worldline $C^{\X}_{(x)}$, but also from all the other
reference worldlines ${C}^{\X}_{(x')}$, $x' \in {\xman}$, of the
original reference frame ${C}^{\X}$.

The correlations $g$ and $p$ between the original metric $\G$ and the original time foliation ${\Sigma}^{\T}$ and reference frame $C^{\X}$ are preserved, in the sense that the transformed metric ${\G}+{\delta}{\G}$ and the
transformed time foliation ${\Sigma}^{{\T}+{\delta}{\T}}$ and
reference frame $C^{{\X}+\delta{\X}}$ yield exactly the same
correlations $g$ and $p$. Crucially, the compatibility conditions
encoded in the set of virtual fields of the theory are respected.
Indeed, since the fields $g$, $N$ and $N^i$ are left invariant by
spacetime diffeomorphisms, (\ref{ychanges 5})-(\ref{ychanges 7}), a
positive definite $g$ remains positive definite, or a positive $N$
remains positive.

\subsection{Infinitesimal symmetries induced by ${\rm Diff}{\xman}$}

These symmetries are generated by vector fields $u^i$ on $\xman$,
and are defined by \begin{eqnarray} \delta y^{\alpha}  &:=&  0  \, ,
\label{xchanges 01}
\\
\delta x^i &:=& u^i        \, , \label{xchanges 02}
\\
\delta t &:=& 0        \, . \label{xchanges 03}
\end{eqnarray}
The fields on ${\rman} \times {\xman}$ respond to these changes via
their Lie derivative ${\cal L}_{u}$ on $\xman$,
\begin{eqnarray}
\delta Y^{\alpha} &=& - {\cal L}_{u}Y^{\alpha} = - Y^{\alpha}_{,i}
u^i \, , \label{xchanges 1}
\\
\delta g_{ij} &=& - {\cal L}_{u}g_{ij}= - g_{ij,k}u^k -
g_{ik}u^k_{,j} -g_{kj}u^k_{,i}    \, , \label{xchanges 2}
\\
\delta N &=& - {\cal L}_{u}N = - N_{,i} u^i \, , \label{xchanges 3}
\\
\delta N^{i} &=& - {\cal L}_{u}N^i = - N^i_{,j} u^j + N^j u^i_{,j}\,
, \label{xchanges 4}
\\
\delta p^{ij} &=& - {\cal L}_{u} p^{ij} =  - p^{ij}_{,k} u^k +
p^{ik}u^j_{,k} + p^{kj}u^i_{,k} - p^{ij} u^k_{,k} \, ,
\label{xchanges 5}
\\
\delta P_{\alpha} &=& 0 \, , \label{xchanges 6}
\\
\delta \Lambda^{\alpha} &=& - {\cal L}_{u} {\Lambda}^{\alpha} =  -
{\Lambda}^{\alpha}_{,k} u^k  \, , \label{xchanges 77}
\end{eqnarray}
where it should be recalled that the momenta $p^{ij}$ are densities
of weight one on $\xman$. The responses of the connecting mappings
$\T$ and $\X$ are determined by the conditions ${\T}(Y)={\rm
Id}_{\rman}$ and ${\X}(Y)={\rm Id}_{\xman}$, where ${\rm
Id}_{\rman}$ and ${\rm Id}_{\xman}$ are the identity transformations
of $\rman$ and $\xman$. Infinitesimally, these relations imply that
\begin{eqnarray}
{\delta}{\T} &=& -{\T}_{,\alpha} \delta{Y}^{\alpha}({\T},{\X}) \, ,
\label{YXT2}
\\
{\delta}{\X}^i &=& -{\X}^i_{,\alpha} \delta{Y}^{\alpha}({\T},{\X})
\, , \label{YXT3}
\end{eqnarray}
which, when combined with (\ref{xchanges 1}), yield
\begin{eqnarray}
\delta {\T} &=& 0 \, , \label{xchanges 7}
\\
\delta {\X}^i &=& u^i({\X}) \, . \label{xchanges 8}
\end{eqnarray}
Using the changes (\ref{xchanges 2})-(\ref{xchanges 4}) and
(\ref{xchanges 7})-(\ref{xchanges 8}) together with the definition
(\ref{G})-(\ref{bfNi}) of $\G$ in terms of $g$, $N$, $N^i$, $\T$ and
$\X$, it follows that $\G$ remains invariant:
\begin{equation}
\delta {\G}_{\alpha \beta} = 0 \, . \label{xchanges 9}
\end{equation}
As discussed in Ref. \cite{kk}, space diffeomorphisms send the
reference worldline $C^{\X}_{(x)}$ of the original reference frame
$C^{\X}$ to the reference worldline
${C}^{{\X}+\delta{\X}}_{(x+{\delta}x)}$ of the same reference frame
$C^{\X}$; i.e., the final reference frame is exactly the same
collection of reference wordlines in $\yman$ as the original. In
addition, space diffeomorphisms do not change the time map $\T$, so
they leave, not only the time foliation ${\Sigma}^{\T}$ but also its
individual instants ${\Sigma}^{\T}_{(t)}$, fixed. While the
spacetime metric field $\G$ is left unchanged, the induced fields on
$\rman \times \xman$ are transformed due to the relabeling of the
wordlines of each reference frame. Regarding the compatibility
conditions, these are respected on account of the fact that each
reference frame and each time foliation is left unchanged, as $\G$
is.

\subsection{Infinitesimal symmetries induced by ${\rm Diff}{\rman}$}

These symmetries are generated by vector fields $w$ on $\rman$, and
are defined by \begin{eqnarray} \delta y^{\alpha}  &:=&  0  \, ,
\label{tchanges 01}
\\
\delta x^i &:=& 0        \, , \label{tchanges 02}
\\
\delta t &:=& w      \, . \label{tchanges 03}
\end{eqnarray}
The fields on ${\rman} \times {\xman}$ respond to these changes via
their Lie derivative ${\cal L}_{w}$ on $\rman$,
\begin{eqnarray}
\delta Y^{\alpha} &=& - {\cal L}_{w}Y^{\alpha} = - Y^{\alpha}_{,t} w
\, , \label{tchanges 1}
\\
\delta g_{ij} &=& - {\cal L}_{w}g_{ij}= - g_{ij,t}w \, ,
\label{tchanges 2}
\\
\delta N &=& - {\cal L}_{w}N = - N_{,t} w - N w_{,t} \, ,
\label{tchanges 3}
\\
\delta N^{i} &=& - {\cal L}_{w}N^i = - N^i_{,t} w - N^i w_{,t} \, ,
\label{tchanges 4}
\\
\delta p^{ij} &=& - {\cal L}_{w} p^{ij} =  - p^{ij}_{,t} w \, ,
\label{tchanges 5}
\\
\delta P_{\alpha} &=& 0 \, , \label{tchanges 6}
\\
\delta {\Lambda}^{\alpha} &=& - {\cal L}_{w} {\Lambda}^{\alpha} =  -
{\Lambda}^{\alpha}_{,t} w - {\Lambda}^{\alpha} w_{,t} \, ,
\label{tchanges 7}
\end{eqnarray}
where it should be recalled that the multipliers $N$, $N^i$ and
$\Lambda$ are densities of weight one on $\rman$. The responses of
the connecting mappings $\T$ and $\X$ are determined using the
relationships (\ref{YXT2})-(\ref{YXT3}) which, when combined with
(\ref{tchanges 1}), yield
\begin{eqnarray}
\delta {\T} &=& w({\T}) \, , \label{tchanges 8}
\\
\delta {\X}^i &=& 0 \, . \label{tchanges 9}
\end{eqnarray}
Using (\ref{tchanges 2})-(\ref{tchanges 4}) and (\ref{tchanges
6})-(\ref{tchanges 9}) together with the definition
(\ref{G})-(\ref{bfNi}) of $\G$ in terms of $g$, $N$, $N^i$, $\T$ and
$\X$, it follows again that $\G$ remains invariant:
\begin{equation}
\delta {\G}_{\alpha \beta} = 0 \, . \label{tchanges 10}
\end{equation}
Regarding the interpretation of these changes \cite{kk}, time
diffeomorphisms send the instant ${\Sigma}^{\T}_{(t)}$ of the
original time foliation ${\Sigma}^{\T}$ onto the instant
${\Sigma}^{{\T}+\delta{\T}}_{(t+\delta{t})}$ of the same time
foliation ${\Sigma}^{\T}$. Although time diffeomorphisms relabel the instants of a given time foliation, the final time foliation consists of exactly the same collection of instants in $\yman$ as the original. In addition, time
diffeomorphisms do not affect the space map $\X$, and hence they
leave, not only the reference frame $C^{\X}$ but also its individual
reference worldlines $C^{\X}_{(x)}$, fixed. While the spacetime
metric field $\G$ is left unchanged, the induced fields on $\rman
\times \xman$ are transformed due to the relabeling of the instants
of each time foliation. The compatibility conditions are again
respected on account of the fact that each reference frame and each
time foliation is left unchanged, as $\G$ is.

\subsection{Infinitesimal symmetries induced by deformations of $\T$ and
$\X$}

The action of space diffeomorphisms and time diffeomorphisms on the manifold points,
\begin{equation}
\delta y = 0 \, , \, \delta t = w(t) \, , \, \delta x^i= u^i(x) \, ,
\label{combined diff1}
\end{equation}
and the connecting mappings,
\begin{equation}
\delta Y = -Y_{,t}w -Y_{,i}u^i \; , \; \delta{\T}=w({\T}) \; , \;
\delta{\X}^i= u^i({\X})  \, , \label{combined diff2}
\end{equation}
can be considered as a single action, which is equivalent to a transformation of the mappings $\T$ and $\X$ while the events $y \in \yman$ and the spacetime metric $\G$ are kept fixed. As we have seen, such a transformation is quite special because it only relabels the hypersurfaces of the time foliation, and the wordlines of the reference frame, without deforming either of them. More general transformations of the mappings $\T$ and $\X$ can now be examined, which generally deform both the time foliation and the reference frame. However, even if such changes formally preserve the field equations derived from the
extended action, they may actually fail to be symmetries of the
formalism. This is because they may violate the restrictions imposed
on the virtual fields $(g, N, N^i)$ of the theory, necessary in
order to ensure compatibility between $\T$, $\X$ and $\G$ at the
level of the solutions.

For example, if changes of the form $\delta{\T}[{\T},{\X}]$ and
$\delta{\X}[{\T},{\X}]$ are considered, which depend on the mappings
$\T$, $\X$ but not on $\G$, then these changes will certainly
violate, at least for some subset of solutions $({\G},{\T},{\X})$,
the compatibility conditions between $\T$, $\X$ and $\G$. Indeed,
given changes $\delta{\T}[{\T},{\X}]$, $\delta{\X}[{\T},{\X}]$ that
deform the time foliation and reference frame while leaving the
spacetime metric unchanged, one can always find a solution $\G$ that
is compatible with the original pair $({\T},{\X})$ but not with the
final pair $({\T}+\delta{\T},{\X}+\delta{\X})$. In this case, the
corresponding final variables $(g+\delta{g}, N+\delta{N},
N^i+\delta{N^i})$ will no longer belong to the set of virtual fields
of the theory; i.e, a positive definite $g$ will no longer be
positive definite, etc.

On the other hand, if the deformations are allowed to depend on all
the fields $\T$, $\X$ and $\G$, then it is indeed possible, at least
in principle, to define symmetries $\delta{\T}[{\T},{\X},{\G}]$ and
$\delta{\X}[{\T},{\X},{\G}]$ of the field equations of the extended
action. The corresponding infinitesimal changes induced on the field
variables are evaluated below. Their finite counterparts are not expected
to form a group; see Pitts and Schieve for a similar phenomenon \cite{PS}. For notational simplicity, the square bracket indicating the functional dependence of the infinitesimal changes $\delta{\T}$ and $\delta{\X}$ on the field variables is not used, but it should be kept in mind that $\delta{\T}$ and $\delta{\X}$ are now functions on both $\rman$ and $\xman$; i.e., they depend on both
$t$ and $x$.

By definition, under these generalised changes $\delta{\T}$,
$\delta{\X}$, we have that
\begin{eqnarray}
\delta y^{\alpha}  &:=&  0  \, , \label{defos 01}
\\
\delta {\G} &:=& 0     \, , \label{defos 02}
\\
\delta x^i &:=& \delta{\X}^i(t,x)  \, , \label{defos 03}
\\
\delta t &:=& \delta{\T}(t,x)   \, . \label{defos 04}
\end{eqnarray}
The fields on ${\rman} \times {\xman}$ respond to these changes via
their Lie derivatives ${\cal L}_{\delta{\T}}$ on $\rman$ and ${\cal
L}_{\delta{\X}}$ on $\xman$ together with some additional terms of
both kinematical and dynamical origin:
\begin{eqnarray}
\delta Y^{\alpha} &=& - {\cal L}_{\delta{\T}}Y^{\alpha} - {\cal
L}_{\delta{\X}}Y^{\alpha} \, , \label{defos 1}
\\
\delta g_{ij} &=& - {\cal L}_{\delta{\T}}g_{ij} - {\cal
L}_{\delta{\X}}g_{ij} -N_i \delta{\T}_{,j}  -N_j \delta{\T}_{,i}  \,
, \label{defos 2}
\\
\delta N &=& - {\cal L}_{\delta{\T}}N - {\cal L}_{\delta{\X}}N + N
N^k \delta{\T}_{,k} \, , \label{defos 3}
\\
\delta N^{i} &=& -{\cal L}_{\delta{\T}}N^i - {\cal
L}_{\delta{\X}}N^i - \delta{\X}^i_{,t} +(N^2 g^{ik}+ N^i N^k)
\delta{\T}_{,k}    \, , \label{defos 4}
\\
\delta {\Lambda}^{\alpha} &=& -{\cal
L}_{\delta{\T}}{\Lambda}^{\alpha} - {\cal
L}_{\delta{\X}}{\Lambda}^{\alpha} - Y^{\alpha}_{,i}
\delta{\X}^i_{,t} \, . \label{defos 5}
\end{eqnarray}

These expressions reproduce the symmetries under ${\rm Diff}{\rman}$
in the special case where $\delta{\T}=w({\T})$ and  $\delta{\X}=0$,
and the symmetries under ${\rm Diff}{\xman}$ in the special case
where $\delta{\T}=0$ and $\delta{\X}=u({\X})$. Next, using the field equations of the covariant canonical action (\ref{maxsym4}), one can determine the changes of the canonical momenta. After some lengthy but straightforward calculation for the metric momenta, one finds that
\begin{eqnarray}
\delta p^{ij} &=& -{\cal L}_{\delta{\T}}p^{ij} - {\cal
L}_{\delta{\X}}p^{ij} + (N^i p^{jk} + N^j p^{ik} - N^k p^{ij})
\delta{\T}_{,k} \nonumber
\\
&& + g^{1\over2} ( 2 g^{ij}g^{kl}- g^{ik}g^{jl}-  g^{il}g^{jk} )
N_{,k} \delta{\T}_{,l} \, , \label{defos 6}
\\
\delta P_{\alpha} &=& 0 . \label{defos 7}
\end{eqnarray}

The interpretation of these Bergmann-Komar \cite{berkom}
deformations is the following: The time foliation ${\Sigma}^{\T}$
and the reference frame $C^{\X}$ are mapped to an entirely different
time foliation ${\Sigma}^{{\T}+{\delta}{\T}}$ and an entirely
different reference frame $C^{{\X}+\delta{\X}}$, while the solution
$\G$ is kept fixed. The compatibility between the unchanged $\G$ and
the transformed mappings ${\T}+{\delta}{\T}$ and ${\X}+{\delta}{\X}$
is preserved only if the functionals $\delta{\T}$ and $\delta{\X}$
generate changes (\ref{defos 2})-(\ref{defos 4}) that preserve the
restrictions imposed on the set of virtual fields of the theory.
Provided that this is the case, the transformed metric $g + \delta
g$ is positive definite, and so on.

\subsection{Infinitesimal symmetries induced by deformations of $Y$}

A symmetrical treatment of the spacetime and the canonical versions
of the extended action suggests that one should also consider
generalised deformations $\delta{Y}[g,p,Y]$ of the foliation
mapping, where the moments $t \in {\rman}$, the points $x \in
{\xman}$ and the fields $g$ and $p$ on ${\rman} \times {\xman}$ are
kept fixed. However, these Bergmann-Komar changes are equivalent to generalised spacetime diffeomorphisms, where a different diffeomorphism is performed in each solution spacetime. Therefore, the infinitesimal changes induced on the field variables of the extended action are identical in form to the ones that have been already considered in the relevant part of this section. The only modification that needs to be made in order to incorporate these
symmetries in the extended framework is to replace the ${\rm
Diff}{\yman}$-induced changes $\delta{Y}=V(Y)$ by the generalised
vector field $\delta{Y}[g,p,Y]$.

\section{The action of symmetries on the phase space}

In this section, the extended equal-time phase space ${\cal
P}=\{g,p,Y,P\}$ is introduced, and the action of symmetries on $\cal
P$ is investigated.

\subsection{Symplectic form and solutions as curves in ${\rman} \times {\cal C}$}

The standard symplectic form on the phase space ${\cal P}$ of the
covariant-canonical action (\ref{maxsym4}) yields the fundamental
Poisson brackets
\begin{eqnarray}
\{ g_{ij}(x), p^{mn}(x') \} &=& {\delta}_{ij}^{mn} \, \delta(x,x')
\, , \label{PB1}
\\
\{ Y^{\alpha}(x), P_{\beta}(x') \} &=& \delta^{\alpha}_{\beta} \,
\delta(x,x') \, , \label{PB2}
\end{eqnarray}
where ${\delta}_{ij}^{mn}:={1 \over 2}({\delta}^m_i {\delta}^n_j +
{\delta}^m_j {\delta}^n_i)$. The Dirac delta function $\delta(x,x')$
is a scalar function on $\xman$ in the first argument and a scalar
density on $\xman$ in the second argument. All the remaining Poisson
brackets between the fields $g$, $p$, $Y$ and $P$ are zero.

Although this is not essential, it is nonetheless clearer to
consider the product space ${\rman}\times{\cal P}$, which is the
natural space where solutions lie. Specifically, each solution
$(g(t),p(t),Y(t),P(t))$ of (\ref{maxsym4}) is viewed as a
one-parameter family $(g_{(t)},p_{(t)},Y_{(t)},P_{(t)})$ of
instantaneous data; that is, as a curve in ${\rman}\times{\cal P}$.
Permissible instantaneous data lie on the constraint surface ${\cal
C} \subset {\cal P}$ of (\ref{maxsym4}) determined by the
super-Hamiltonian and super-momentum constraints (\ref{Ham
con})-(\ref{Mom con}), and the annihilation of the embedding
momenta, (\ref{conB}). The curves representing the solutions of
(\ref{maxsym4}) lie on the subspace ${\rman}\times{\cal C}$ of
${\rman}\times{\cal P}$, and each moment $t_o \in {\rman}$ defines a
copy $t_o \times {\cal C} \in {\rman}\times{\cal C}$ of the
constraint surface $\cal C$.

The primary role of the Hamiltonian
\begin{equation}
\int d^3x \, {\cal H}:=\int d^3x \,
(NH+N^iH_i+{\Lambda}^{\alpha}P_{\alpha}) \label{calH}
\end{equation}
of (\ref{maxsym4}) is to generate the dynamical evolution of the
system. More precisely, $\int d^3x \, {\cal H}$ generates a curve
$(g_{(t)},p_{(t)},Y_{(t)},P_{(t)})$ in ${\rman}\times{\cal C}$ from
an initial datum $(g_{(t_o)},p_{(t_o)},Y_{(t_o)},P_{(t_o)})$ in $t_o
\times {\cal C}$ via its Poisson bracket action
\begin{equation}
f_{(t_o + \delta t)}(x) = f_{(t_o)}(x) +  \delta{f}_{(t_o)}(x) =
{f}_{(t_o)}(x) + \{ {f}(x), \int d^3x \, {\cal H} \} {\big |}_{f =
f_{(t_o)}} \, \delta t \, , \label{infin action}
\end{equation}
where $f$ denotes any of the canonical variables. The symbol
$\{,\}{\big |}_{f= f_{(t_o)}}$ means that the phase-space fields
$(g,p,Y,P)$ must be evaluated at the particular initial datum
$(g_{(t_o)},p_{(t_o)},Y_{(t_o)},P_{(t_o)})$ after the Poisson
brackets have been worked out. The multipliers $N$, $N^i$ and
$\Lambda$ appearing in (\ref{infin action}) are regarded as smearing
functions on $\xman$; i.e, they carry no $t$-dependence at this
stage. Different choices of such functions generate different curves
in ${\rman} \times {\cal C}$.

\subsection{The action of symmetries on solutions}

Assuming that all solutions, i.e., complete dynamical trajectories
of (\ref{maxsym4}), have been generated by $\cal H$ via (\ref{infin
action}), the multipliers $N$, $N^i$ and $\Lambda$ now become
functions of $t$: that is, for each given solution
$(g_{(t)},p_{(t)},Y_{(t)},P_{(t)})$ in ${\rman}\times{\cal C}$, the
multipliers can be in principle evaluated as functions of $t$ using
the field equations. A symmetry of the field equations of
(\ref{maxsym4}) acts on each one of these solutions according to the
expressions derived in section 5. It maps each curve
\begin{equation}
{\Big (} \, g_{(t)} \, , \, p_{(t)} \, , \, Y_{(t)} \, , \, P_{(t)}
\, {\Big )} \in {\rman} \times {\cal C} \, \label{orig curve}
\end{equation}
to a neighbouring curve
\begin{equation}
{\Big (} \, (g+\delta g)_{(t)} \, , \, (p+\delta p)_{(t)} \, , \,
(Y+\delta Y)_{(t)} \, , \, (P+\delta P)_{(t)} \, {\Big )} \in
{\rman}\times{\cal C} \, , \label{final curve}
\end{equation}
where the changes $\delta g_{(t)}$, $\delta p_{(t)}$, $\delta
Y_{(t)}$ and $\delta P_{(t)}$ depend in general on the particular
curve and symmetry considered. The actions of the diffeomorphisms of
$\yman$, $\xman$ and $\rman$ and the action of the generalised
deformations $\delta{\T}$, $\delta{\X}$ and $\delta{Y}$ on these
curves have been already investigated in section 5. In this section,
the dynamical variables in ${\cal P}$ that generate these actions
are identified.

\subsection{The canonical generator of ${\rm Diff}{\yman}$}

For convenience, let us summarise the changes induced by ${\rm
Diff}{\yman}$ on each solution curve in ${\rman} \times {\cal C}$.
Copying these changes from section 5, we have:
\begin{eqnarray}
\delta g_{ij} &=& 0 \, , \nonumber
\\
\delta p^{ij} &=& 0 \, , \nonumber
\\
\delta Y^{\alpha} &=& V^{\alpha}(Y) \, , \nonumber
\\
\delta P_{\alpha} &=& - P_{{\beta}} V^{\beta}_{,\alpha}(Y) \, .
\nonumber
\end{eqnarray}
If $f_{(t)}(x)$ denotes any of these solution components, the
general change $\delta f_{(t)}(x)$ in ${\rman} \times {\cal C}$ is
weakly reproduced via the Poisson brackets
\begin{equation}
\delta f_{(t)}(x) = \{ f(x) , {\cal D}_{V} \}{\big |}_{f(x) =
f_{(t)}(x)} \, \label{can diffm}
\end{equation}
of $f(x)$ with the dynamical variable
\begin{equation}
{\cal D}_{V} = \int d^3x P_{\alpha}(x) V^{\alpha}(Y(x)) \, .
\label{can diffm2}
\end{equation}
The generator ${\cal D}_{V}$ depends solely on the embedding
variables and the generating vector field $V$, and therefore has no
counterpart in the Dirac-ADM theory. Moreover, the ultra-locality of
${\cal D}_{V}$ in $t$; i.e., the fact that ${\cal D}_{V}$ does not
contain any $t$-derivatives, implies that the action of this
generator on ${\rman} \times {\cal C}$ depends only on the points $c
\in {\cal C}$. Therefore, this action induces a well-defined action
on $\cal C$, and spacetime diffeomorphisms can be represented by
symplectic diffeomorphisms of the phase space. Specifically, the
generator ${\cal D}_{V}$ defines an anti-homomorphic mapping from
vector fields in the Lie algebra of ${\rm Diff}{\yman}$ into the
Poisson bracket algebra on the phase space of the system:
\begin{equation}
\{ {\cal D}_{V_1} , {\cal D}_{V_2} \} = {\cal D}_{-[V_1,V_2]} \, ,
\label{diffmrep}
\end{equation}
where the Lie bracket $[V_1,V_2]$ of the vector fields $V_1$ and
$V_2$ is defined by
\begin{equation}
[V_1,V_2]^{\alpha}:=V_1^{\beta}V_{2,\beta}^{\alpha} -
V_2^{\beta}V_{1,\beta}^{\alpha} \, . \label{diffmrep2}
\end{equation}
The reason that this representation is established via an
anti-homomorphism, rather than a homomorphism, has been discussed
extensively in Refs. \cite{ik} and \cite{kk}.

\subsection{The canonical generator of ${\rm Diff}{\xman}$}

The changes induced by ${\rm Diff}{\xman}$ on each solution curve in
${\rman} \times {\cal C}$ are:
\begin{eqnarray}
\delta g_{ij} &=& - g_{ij,k} u^k - g_{ik}u^k_{,j} -g_{kj} u^k_{,i}
\, , \nonumber
\\
\delta p^{ij} &=& - p^{ij}_{,k} u^k + p^{ik}u^j_{,k} +
p^{kj}u^i_{,k} - p^{ij} u^k_{,k} \, , \nonumber
\\
\delta Y^{\alpha} &=& - Y^{\alpha}_{,i} u^i \, , \nonumber
\\
\delta P_{{\alpha}} &=& - P_{{\alpha},k} u^k - P_{\alpha} u^k_{,k}
\, . \nonumber
\end{eqnarray}
Letting again $f_{(t)}(x)$ denote any of these solution components,
the general change $\delta f_{(t)}(x)$ in ${\rman} \times {\cal C}$
is weakly reproduced via the Poisson brackets
\begin{equation}
\delta f_{(t)}(x) = \{ f(x) , {\cal D}_{u} \}{\big |}_{f(x) =
f_{(t)}(x)} \, \label{can diffx}
\end{equation}
of $f(x)$ with the dynamical variable
\begin{equation}
{\cal D}_{u} = \int d^3x \, u^i(x) \, {\Big(} H_i(x) + P_{\alpha}(x)
Y^{\alpha}_{,i}(x) {\Big)} \, . \label{can diffx2}
\end{equation}
The generator ${\cal D}_{u}$ is ultralocal in $t$, which again
provides the key to representing space diffeomorphism by symplectic
diffeomorphisms of $\cal P$. Specifically, a homomorphism arises,
\begin{equation}
\{ {\cal D}_{u_1} , {\cal D}_{u_2} \} = {\cal D}_{[u_1,u_2]} \, ,
\label{diffxrep}
\end{equation}
where the Lie bracket $[u_1,u_2]$ of the vector fields $u_1$ and
$u_2$ on $\xman$ is given by
\begin{equation}
[u_1,u_2]^{i}:=u_1^{j}u_{2,j}^{i} - u_2^{j}u_{1,j}^{i} \, .
\label{diffxrep2}
\end{equation}

\subsection{The canonical generator of ${\rm Diff}{\rman}$}

The changes induced by ${\rm Diff}{\rman}$ on each solution curve in
${\rman} \times {\cal C}$ are:
\begin{eqnarray}
\delta g_{ij} &=& - g_{ij,t} w \, , \nonumber
\\
\delta p^{ij} &=& - p^{ij}_{,t} w \, , \nonumber
\\
\delta Y^{\alpha} &=& - Y^{\alpha}_{,t} w \, , \nonumber
\\
\delta P_{{\alpha}} &=& - P_{{\alpha},t} w \, . \nonumber
\end{eqnarray}
The general change $\delta f_{(t)}(x)$ in ${\rman} \times {\cal C}$
induced by time diffeomorphisms is again weakly reproduced via the
Poisson brackets
\begin{equation}
\delta f_{(t)}(x) = \{ f(x) , {\cal D}_{w} \}{\big |}_{{f}(x) =
f_{(t)}(x),} \, \label{can difft}
\end{equation}
of $f(x)$ with the dynamical variable
\begin{equation}
{\cal D}_{w} = - w(t) \int d^3x \, {\cal H}(x) \, , \label{can
difft2}
\end{equation}
where $\int d^3x \, {\cal H}$ is the Hamiltonian (\ref{calH}).
However, the smearing functions $N(x)$, $N^i(x)$ and $\Lambda(x)$
that are present in (\ref{can difft2}) must be replaced at the end
of the calculation by the lapse $N_{(t)}(x)$, shift $N_{(t)}^i(x)$
and multipliers ${\Lambda}_{(t)}(x)$ associated with the particular
solution curve in ${\rman} \times {\cal C}$ whose change is
considered. This means that the action of the generator ${\cal
D}_{w}$ on ${\rman} \times {\cal C}$ depends not only on the point
$c \in {\cal C}$, but also on the particular curve passing through
this point. Hence, it does not induce a well-defined action on $\cal
C$. In this case, the Lie algebra of vector fields on ${\rman}$
cannot be represented by symplectic diffeomorphisms of $\cal P$.
This is possible only within a history phase space formulation of
the canonical action (\ref{maxsym4}).

\subsection{The canonical generator of generalised deformations $\delta{\T}$, $\delta{\X}$}

The changes induced on each solution curve in ${\rman} \times {\cal
C}$ by the deformations $\delta{\T}$ and $\delta{\X}$ of the
mappings are:
\begin{eqnarray}
\delta g_{ij} &=& - {\cal L}_{\delta{\T}}g_{ij} - {\cal
L}_{\delta{\X}}g_{ij} -N_i \delta{\T}_{,j}  -N_j \delta{\T}_{,i}  \,
, \nonumber
\\
\delta p^{ij} &=& -{\cal L}_{\delta{\T}}p^{ij} - {\cal
L}_{\delta{\X}}p^{ij} + (N^i p^{jk} + N^j p^{ik} - N^k p^{ij})
\delta{\T}_{,k} \nonumber
\\
&& + g^{1\over2} ( 2 g^{ij}g^{kl}- g^{ik}g^{jl}-  g^{il}g^{jk} )
N_{,k} \delta{\T}_{,l} \, , \nonumber
\\
\delta Y^{\alpha} &=& - {\cal L}_{\delta{\T}}Y^{\alpha} - {\cal
L}_{\delta{\X}}Y^{\alpha} \, , \nonumber
\\
\delta P_{{\alpha}} &=& 0 \, . \nonumber
\end{eqnarray}
The general change $\delta f_{(t)}(x)$ in ${\rman} \times {\cal C}$
induced by time diffeomorphisms is weakly reproduced via the Poisson
brackets
\begin{equation}
\delta f_{(t)}(x) = \{ f(x) , {\cal D}_{({\delta{\T}},{\delta{\X}})}
\}{\big |}_{{f}(x) = f_{(t)}(x),} \, \label{can difft3}
\end{equation}
of $f(x)$ with the dynamical variable
\begin{equation}
{\cal D}_{({\delta{\T}},{\delta{\X}})} = - \int d^3x \, \Big( \,
\delta{\T}(NH+N^iH_i +{\Lambda}^{\alpha}P_{\alpha}) + \delta{\X}^i
(H_i + P_{\alpha}Y^{\alpha}_{,i}  ) \, \Big) \, . \label{can defos}
\end{equation}
Again, the smearing functions $N(x)$, $N^i(x)$ and $\Lambda(x)$ that
are present in (\ref{can defos}) must be replaced at the end of the
calculation by the lapse $N_{(t)}(x)$, shift $N_{(t)}^i(x)$ and
multipliers ${\Lambda}_{(t)}(x)$ associated with the particular
solution curve in ${\rman} \times {\cal C}$ whose change is
considered. Hence, it does not induce a well-defined action on $\cal C$.
The representation of such deformations is again possible only within a history phase space formulation, an issue discussed in detail in Ref.
\cite{s2}. Finally, it should be noted that the generator ${\cal
D}_{({\delta{\T}},{\delta{\X}})}$ reduces to the generator ${\cal
D}_{w}$ of time diffeomorphisms if $\delta{\T}=w({\T})$ and
$\delta{\X}=0$, and to the generator ${\cal D}_{u}$ of space
diffeomorphisms if $\delta{\T}=0$ and $\delta{\X}=u({\X})$.

\subsection{The canonical generator of generalised deformations $\delta{Y}$}
As discussed in the previous section, such deformations correspond
to generalised spacetime diffeomorphisms. The general change $\delta
f_{(t)}(x)$ in ${\rman} \times {\cal C}$ of any solution component
$f_{(t)}(x)$ is weakly reproduced via the Poisson brackets
\begin{equation}
\delta f_{(t)}(x) = \{ f(x) , {\cal D}_{\delta{Y}} \}{\big |}_{f(x)
= f_{(t)}(x)} \, \label{can diffm35}
\end{equation}
of $f(x)$ with the dynamical variable
\begin{equation}
{\cal D}_{\delta{Y}} = \int d^3x P_{\alpha}(x) \delta{Y}^{\alpha}(x)
\, . 
\label{can diffm45}
\end{equation}
Of course, since $\delta{Y}^{\alpha}$ now depends upon all the canonical variables, it does not weakly commute with $g$ and $p$. Therefore, in order for (\ref{can diffm35})-(\ref{can diffm45}) to reproduce the correct action on $g$ and $p$, explicit use of the constraint $P_{\alpha}=0$ must be made.

\section{Observables and their evolution}

An inspection of the canonical transformations of the previous
section reveals two facts about canonical general relativity that
are perhaps unexpected. First, although it is true that the
invariance of the spacetime action under ${\rm Diff}{\yman}$ implies
four first-class constraints, these are not the super-Hamiltonian
and the super-momentum constraints but rather the embedding momentum
constraints. Second, the orbits of the generator of spacetime
diffeomorphism ${\cal D}_{\yman}$ in the phase space are distinct
from the orbits of the Hamiltonian $\int  d^3x \, {\cal H}$. This
eliminates any possibility of regarding the Hamiltonian as the
generator of spacetime diffeomorphisms, in agreement with the
long-standing viewpoint of Kucha\v{r} \cite{kuch92}.

These results are analogous to the results obtained by Savvidou in
the context of the history formulation of general relativity
\cite{s1}-\cite{s3} and those obtained in Ref. \cite{kk} for the
Bosonic string. In particular, the distinction between the
gravitational Hamiltonian $\int  d^3x \, {\cal H}$ and the generator
of diffeomorphisms ${\cal D}_{\yman}$ reflects analogous
distinctions in \cite{kk}-\cite{s3}. This has consequences for the observables of the theory and their dynamical evolution, a fact that, within the history framework, was pointed out in \cite{s2}-\cite{s3}.

\subsection{Spacetime and Dirac observables}

In the present context, functions on the sets $\Gamma$ and $\Delta$
of (\ref{maxsym4}) are connected with two kinds of instantaneous
observables: spacetime observables and Dirac observables. In
general, two functions $F_1$ and $F_2$ are defined to be equivalent
on the phase space $\cal P$ of (\ref{maxsym4}) if their values are
equal on the constraint surface ${\cal C} \subset {\cal P}$,
\begin{equation}
F_1{\mid}_{\cal C} = F_2{\mid}_{\cal C} \, . \label{equivalent
functions2}
\end{equation}
The surface $\cal C$ is determined by the constraints (\ref{Ham
con})-(\ref{Mom con}) and (\ref{conB}). An equivalence class of such
functions, denoted by $F$, will be called a {\it spacetime
observable} if it is invariant under ${\rm Diff}{\yman}$; i.e., if
$F$ commutes on $\cal C$ with the generator ${\cal D}_{V}$,
\begin{equation}
\{ F , {\cal D}_{V} \}{\mid}_{\cal C} = 0 \, ,
\label{SPTM-observables}
\end{equation}
where ${\cal D}_{V}$ is given by (\ref{can diffm2}). From the
elementary functions $F^{g}:=g$, $F^{p}:=p$, $F^{Y}:=Y$ and
$F^{P}:=P$ on $\cal P$, only the first three are non-trivial,
because $F^{P}$ is equivalent to the zero function. From these,
$F^{g}$ and $F^{p}$ are spacetime observables, \begin{eqnarray} \{
F^{g} , {\cal D}_{V} \}{\mid}_{\cal C} = 0 \, , \label{inv of g}
\\
\{ F^{p} , {\cal D}_{V} \}{\mid}_{\cal C} = 0 \, , \label{inv of p}
\end{eqnarray}
while $F^{Y}=Y$ is not a spacetime observable,
\begin{equation}
\{ F^{Y} , {\cal D}_{V} \}{\mid}_{\cal C} = V(Y) \, . \label{noninv
of Y}
\end{equation}
In general, any dynamical variable on $\cal P$ that is independent
of $Y$ is a spacetime observable. The {\it Dirac observables} form a
subset of spacetime observables that are additionally invariant
under the action of the generator ${\cal
D}_{({\delta{\T}},{\delta}{\X})}$ of deformations. An equivalent
description of the Dirac observables, which follows the standard
definition, is that $F^{\rm Dir}$ is a Dirac observable if it
commutes on $\cal C$ with all first-class constraints:
\begin{eqnarray}
\{ F^{\rm Dir} , P_{\alpha} \}{\mid}_{\cal C} &=& 0 \, ,
\label{Dirac obs 1}
\\
\{ F^{\rm Dir} , H \}{\mid}_{\cal C} &=& 0 \, , \label{Dirac obs 2}
\\
\{ F^{\rm Dir} , H_i \}{\mid}_{\cal C} &=& 0 \, . \label{Dirac obs
3}
\end{eqnarray}
Any such function also commutes with the Hamiltonian of
(\ref{maxsym4}).

\subsection{The evolution of spacetime observables}

Each spacetime observable $F(g,p,Y,P)$ represents the ${\rm
Diff}{\yman}$-invariant result of measuring an instantaneous
canonical state of the system. Given any solution curve
$(g_{(t)},p_{(t)},Y_{(t)},P_{(t)})$ in ${\rman} \times {\cal C}$,
the dynamical evolution of a spacetime observable $F$ can be defined
by comparing the values of $F$ at various times $t$ along this
particular solution
\begin{equation}
F_{(t)}(g,p,Y,P) := F(g_{(t)},p_{(t)},Y_{(t)},P_{(t)}) \, .
\label{evol of obs}
\end{equation}
In other words, one compares the results of the {\it same}
measurement at different times. Particular attention should be paid
to the fact that $t$ is a ${\rm Diff}{\yman}$-invariant structure,
according to equation (\ref{ychanges 03}). The evolution of the
elementary spacetime observables $F^{g}:=g$ and $F^{p}:=p$ is given
by
\begin{eqnarray}
F^{g}_{(t)}(g,p,Y,P) &=& F^{g}(g_{(t)},p_{(t)},Y_{(t)},P_{(t)}) =
g_{(t)} \, , \label{evol of obs 1}
\\
F^{p}_{(t)}(g,p,Y,P) &=& F^{p}(g_{(t)},p_{(t)},Y_{(t)},P_{(t)}) =
p_{(t)} \, . \label{evol of obs 2}
\end{eqnarray}
Each Dirac observable $F^{\rm Dir}$ is a spacetime observable that does not
evolve in time,
\begin{equation}
F^{\rm Dir}_{(t)}(g,p,Y,P) = F^{\rm
Dir}(g_{(t)},p_{(t)},Y_{(t)},P_{(t)}) =  F^{\rm Dir}(g,p,Y,P) \, ,
\label{evol of obs 3}
\end{equation}
on account of equations (\ref{Dirac obs 1})-(\ref{Dirac obs 3}).

\subsection{The relation between observables and the sets $\Gamma$ and $\Delta$}

The connection between the spacetime observables, their subset of
Dirac observables, and the sets $\Gamma$ and $\Delta$ of the
extended action is as follows. Each element of the set of solutions
of (\ref{maxsym4}) defines a curve
$(g_{(t)},p_{(t)},Y_{(t)},P_{(t)})$ in ${\rman} \times {\cal C}$. As
discussed in section 5, any two solutions in this set that can be
mapped onto each other by a spacetime diffeomorphism must have
identical components $g_{(t)}$ and $p_{(t)}$ and differ only in
their component $Y_{(t)}$. Hence, each element $\gamma$ of the set
$\Gamma$ of ${\rm Diff}{\yman}$-classes of solutions of
(\ref{maxsym4}) is characterised by the solution-components
$g_{(t)}$ and $p_{(t)}$. Therefore, in order to be able to
distinguish between the elements of $\Gamma$ based on instantaneous
measurements of the system, one has to measure the instantaneous
data $g$ and $p$ at all times $t$. The resulting collection
$(g_{(t)},p_{(t)})$, $t \in {\rman}$ of values then determines a
unique element of $\Gamma$. Accordingly, the collection $\{ (
F^{g}_{(t)}, F^{p}_{(t)}) \}$, $t \in {\rman}$, of the elementary
spacetime observables is able to distinguish between the elements of
$\Gamma$. Such a collection can be regarded as inducing a complete
set of functions on $\Gamma$.

Referring to section 5 again, the set $\Delta$ of first-class orbits
of (\ref{maxsym4}) is derived from the set $\Gamma$ by identifying
two elements ${\gamma}_1=(g_1,p_1)$ and ${\gamma}_{2}=(g_2,p_2)$ of
$\Gamma$ that reconstruct via arbitrary choices of the foliation
mapping $Y$ the same ${\rm Diff}{\yman}$-class $\{ \G \}$ of
spacetime solutions. In the spacetime representation, ${\gamma}_1$
and ${\gamma}_{2}$ are denoted by ${\rm Diff}{\yman}$-classes of
solutions; i.e., ${\gamma}_1=\{( {\G}_{1}, {\T}_{1}, {\X}_{1}  ) \}$
and ${\gamma}_2=\{( {\G}_{2}, {\T}_{2}, {\X}_{2} ) \}$. Two such
elements of $\Gamma$ define the same element $\delta \in \Delta$ if
the metric classes $\{ {\G}_1 \}$ and $\{ {\G}_2 \}$ coincide. The
mappings $\T$ and $\X$ are not involved at all in this last
consideration. As a result, two elements $s_1 =
({\G}_{1},{\T}_{1},{\X}_{1})$ and $s_2 =
({\G}_{2},{\T}_{2},{\X}_{2})$ of the set of solutions of
(\ref{maxsym4}) define the same element $\delta \in \Delta$ if they
can be mapped onto each other by a combined action of spacetime
diffeomorphisms and deformations of the mappings $\T$ and $\X$. In
phase space, this corresponds to a combined action of the generators
${\cal D}_{\yman}$ and ${\cal D}_{({\delta{\T}},{\delta{\X}})}$. Any
function on $\cal P$ that is invariant under the action of both
these generators, commutes on $\cal C$ with all eight first-class
constraints, and hence projects down to a function on $\Delta$.
These considerations reexpress the standard result that Dirac
observables induce functions on $\Delta$.

\subsection{On the problem of time}

As emphasised in the introduction, the covariant canonical formalism
admits two interpretations. The conventional interpretation, of
Bergmann and Dirac, treats all first-class constraints as gauge
generators. This implies that the diffeomorphisms of ${\yman}$ and
the deformations of the mappings $\T$ and $\X$ have no measurable
consequences. The fields $\T$ and $\X$ are regarded as physically
unimportant, and only the subset $\Delta \subset \Gamma$ retains its
physical significance. In this case, the fact that $t$ is a {\rm
Diff}{\yman}-invariant parameter, in accordance with equation
(\ref{ychanges 03}), is not sufficient to guarantee that it is a
gauge-invariant parameter. This is because, in accordance with
equation (\ref{tchanges 03}), $t$ is not invariant under a
deformation of the mapping $\T$, which is also treated as a gauge
transformation. Thus, the term observable is reserved for the Dirac
observables, and the classical problem of time re-appears in its
standard form. If this interpretation is accepted, the only
surviving attribute of the extended action is its ability to
accommodate the representations of the Lie algebra of ${\rm
Diff}{\yman}$.

Actually, it is not only the Lie algebra, but the group itself, that
can be represented within the phase space of (\ref{maxsym4}). In
particular, a conceptual difficulty which arose in the formalism of
Isham and Kucha\v{r} does not arise here. This difficulty, analysed
in Ref. \cite{ik}, is that the space of spacelike embeddings is not
an invariant submanifold of the space of all embeddings. Hence, the
vector fields on the space of spacelike embeddings are not complete,
and only the Lie algebra of spacetime diffeomorphisms can be
represented in the phase space of \cite{ik}. This is not a problem
here on account of the fact that the compatibility conditions are
not imposed directly on the foliation variable $Y$. Instead, they
are incorporated into the definition of the virtual fields
$(g,N,N^i)$. Since the metric $\G$ and the mappings $\T$ and $\X$
all transform covariantly under spacetime diffeomorphisms, the
fields $g$, $N$ and $N^i$ remain invariant under these
diffeomorphisms, and the compatibility conditions incorporated in
them are preserved. The fact that the vector fields $\delta{Y}$ do
not need to be restricted is especially evident at the level of the
variation of the action (\ref{maxsym4}), which leaves the mapping
$Y$ entirely undetermined.

This property of the covariant canonical formalism also solves the
problem of interpreting the so-called microcausality condition in
canonical quantum gravity. The microcausality condition, discussed,
for example, by Isham in Ref. \cite{time1}, is the requirement that
quantum operators should commute for all spacetime points that are
spacelike separated. As originally realised by Fredenhagen and Haag
\cite{FredHag}, for most pairs of points in spacetime there exists
at least one Lorentzian metric with respect to which these points
are not spacelike separated. Insofar as all metrics are virtually
present in quantum theory, the microcausality consition is violated.
However, as argued  in Refs. \cite{s2}-\cite{s3}, this problem is
overcome provided that both the foliation and the spacetime metric
transform covariantly, which is also the case here.

These points acquire physical significance once the alternative, and richer, interpretation of the formalism is accepted. Then, the entire set $\Gamma$ is considered physically meaningful, and the theory retains its full effect. The observable aspects of vacuum gravity are not limited to the Dirac
observables, and spacetime observables arise whose dynamical
evolution is non-trivial. In particular, for each physical state
$\delta \in \Delta$, there corresponds a whole set $\{ \gamma \}$ of
states $\gamma \in \Gamma$, all of which are foliation-dependent but
${\rm Diff}{\yman}$-invariant and, hence, observable. Kucha\v{r}'s
terminology is then quite appropriate: Dirac observables were
referred to as perennials in Ref. \cite{kuch92}, precisely in order
to separate them from other observable aspects of general relativity
that evolve in time. Kucha\v{r} criticised the identification of the
Hamiltonian with a gauge generator, and proposed certain evolving
observables that do not weakly commute with it.

The spacetime observables derived from (\ref{maxsym4}) represent
such observable aspects of the vacuum theory. In particular, the
elementary functions $F^g=g$ and $F^p=p$ on the phase space $\cal P$
are the prototypes of evolving, foliation-dependent, spacetime
observables. Kucha\v{r}'s proposal for observables in Ref.
\cite{kuch92} does not entirely coincide with the present one, but
corresponds to an intermediate viewpoint that regards the
deformations of $\X$ as gauge, but those of $\T$ as measurable.
According to this standpoint, the elementary functions $F^g=g$ and
$F^p=p$ are not observables, but any ${\rm Diff}{\xman}$-invariant
functional of $g$ and $p$ is. Of course, accepting here $\X$ as a
gauge field would imply an asymmetrical treatment of the mappings
$\T$ and $\X$, and this would not be supported by the assignment of
a reference system to each pair $({\T},{\X})$.

Finally, a comment can be made about the ${\rm
Diff}{\yman}$-invariant time parameter $t$, upon which the
interpretation of the covariant canonical formalism rests. Although
the energy-momentum of the fields ${\T}$ and ${\X}$ vanishes in
(\ref{maxsym}), and hence the time parameter $t$ of (\ref{maxsym4})
cannot be regarded as being part of the system in the strict sense,
this may not be such an undesired property after all. For this
notion of time is not very distant from the notion of time arising
in conventional quantum field theory---or in histories theory \cite{s1}-\cite{s3} as the parameter of partial ordering. There, too, time is external to the system, in the sense that it forms part of the background on which the system evolves, in accordance with the Copenhagen interpretation. Actually, the parameter $t$ of (\ref{maxsym4}) is as external as a time can ever be in a closed system such as general relativity. This may be regarded as an indication that the covariant canonical action provides a link between the special and the general theory of relativity or, more generally, between a theory on a given background and gravitation theory. However, approaching quantum
gravity in this light presupposes a quantisation method which, at
the very least, should respect the distinction between the sets
$\Gamma$ and $\Delta$. If such a method is attainable remains to be
studied.

\section{Acknowledgments}

I would like to thank Charis Anastopoulos, Petr H\'{a}j\'{\i}\v{c}ek, Chris
Isham, Karel Kucha\v{r} and Ntina Savvidou for reading the final draft of
this paper.

\bibliographystyle{unsrt}

\end{document}